\documentclass[review]{elsarticle}

\usepackage{geometry}
\usepackage{float}
\usepackage{amsfonts}
\usepackage{amsmath}
\usepackage{cases}
\usepackage{color}
\usepackage{subfig} %子图扩展包(subfloat)， added by 李志斌
\renewcommand{\b}[1]{{\boldsymbol{#1}}}
\renewcommand{\r}[1]{{\mathrm{#1}}}

\newtheorem{remark}{Remark}

\journal{arXiv}

\begin{document}

\begin{frontmatter}

\title{A peridynamics-based finite element method for quasi-static fracture analysis}

\author[a]{Fei Han\corref{cor}}
\cortext[cor]{Corresponding author}
\ead{hanfei@dlut.edu.cn}

\author[a]{Zhibin Li}

\address[a]{State Key Laboratory of Structural Analysis for Industrial Equipment, Department of Engineering Mechanics, Dalian University of Technology, Dalian 116023, China}

\begin{abstract}
In this paper, a peridynamics-based finite element method (Peri-FEM) is proposed for the quasi-static fracture analysis, which is of the consistent computational framework with the classical finite element method (FEM). First, the integral domain of the peridynamics is reconstructed, and a new type of element called peridynamic element (PE) is defined. Although PEs are generated by the continuous elements (CEs) of classical FEM, they do not affect each other. Then the spatial discretization is performed based on PEs and CEs, and the linear equations about the nodal displacement are established according to the principle of minimum potential energy. Besides, the cracks are characterized as the degradation of the mechanical properties of PEs. Finally, the validity of the proposed method is demonstrated through numerical examples.
\end{abstract}


\end{frontmatter}

%---------------------1---------------------%
\section{Introduction}
The prediction of the fracture phenomena for the structures is of great significance in engineering practice. Although, for such issues, the classical finite element method (FEM) based on the classical continuum mechanics seems powerless. Various improved techniques have been proposed to deal with the fracture properly. Either the mathematical model describing the structural deformation or the numerical method solving the mathematical model is improved.

The extended finite element method (XFEM) is an upgraded version of the classical FEM \cite{10belytschko1999elastic,12daux2000arbitrary}. By improving the shape function of FEM, i.e., introducing the enrichment functions, this method can be used to analyze the discontinuity problem without remeshing \cite{11moes1999finite}. However, the crack geometries need to be tracked to determine the enrichments nodes in this method. Thus, although the XFEM approximation can represent crack geometries independent of element boundaries, it also relies on the interaction between the mesh and the crack geometry to determine the sets of enriched nodes \cite{13bellec2003note}. In addition, the XFEM is not convenient in dealing with complex cracking, such as branching, due to the need to track cracks. 

Such difficulties can be alleviated by improving the mathematical model for fracture analysis. The phase field approach for fracture is such a model. It originates from the variational formulation of brittle fracture \cite{14francfort1998revisiting} and its regularization version \cite{15bourdin2000numerical}. The phase field model avoids tracking the complicated crack geometries; instead, the phase field variable representing material degradation is introduced to describe the crack evolution. It then leads to a coupled problem and can be solved with the classical FEM. However, this method uses the damage zone to represent the crack and cannot model the discontinuous surface. Besides, to resolve high gradients of the phase field appearing in the transition zones between cracked and uncracked materials, the regularization parameter controlling the width of the damage zone must be smaller enough \cite{16ziaei2016massive}, which means the mesh must be fine enough. Thus the computational cost is very expensive.

The peridynamics is a nonlocal theory proposed by Silling et al \cite{1silling2000reformulation, 17silling2007peridynamic}. The equation of motion for the peridynamics is an integral-differential equation, which does not contain the spatial derivation, thus can allow the space discontinuities naturally. In the peridynamics, non-contact material points in the object interact with each other through the bond. Then, the breaking of a bond, which can be determined according to the bond deformation, is used to describe cracking. Thereby the issues of complicated crack tracking can be avoided. Therefore, the peridynamics is a promising model for solid mechanics. However, the analytical solution of the integral-differential equation for the peridynamics is usually challenging to obtain. Thus, its numerical implementation technologies are essential.

The most commonly used numerical strategy for the peridynamics is the mesh-free method \cite{6silling2005meshfree}. In this method, the reference configuration is discretized into nodes with a known volume, then the integral in the equation of motion is replaced by a finite sum. Thus, the mesh-free strategy can capture the crack geometry freely. However, the body needs to be discretized with a large number of nodes to ensure calculation accuracy. Besides, the accuracy of computation decays dramatically in the case of non-uniform discretization \cite{20ren20173d}. An alternative way for the numerical implementation of the peridynamics is the discontinuous Galerkin finite element method (DGFEM) \cite{20ren20173d,19chen2011continuous,5azdoud2014morphing}. Although the technical elements of this method, such as the shape function, are consistent with the classical FEM based on the classical continuum mechanics, the computational framework is different and more troublesome. This is because the total potential energy in the peridynamic theory contains a double integral term.

For the numerical implementation of the peridynamics, the mesh-free method is adopted by most researchers. However, a few studies to implement the peridynamics by FEM or DGFEM has not received much attention, which may result from the inconsistency of the implementation process from the classical FEM. 

In this paper, the integral domain of the peridynamics is reconstructed to transform the double integral to the single integral in the total potential energy. Then a new type of element, called peridynamic element (PE), is defined in the new integral domain to keep the properties of single integral in discrete form. The PEs are constructed based on the elements in classical FEM, which are characterized as sharing nodes between adjacent elements, thus called continuous element (CE) in this paper. The PEs do not have any special requirements for the type of CEs. With the PE meshes at hand, the element-based discrete numerical method for the peridynamics is reorganized, and a computational framework consistent with classical FEM is obtained. Besides, the original way of using element separation to characterize cracks is converted into using the degradation of PE's mechanical properties. In this way, the degree of solution freedom can be reduced dramatically with the same number of elements. It is worth pointing out that although PEs are defined based on CEs, they do not affect each other. 

The paper is organized as follows. The  basic formulations and the principle of the minimum potential energy for the bond-based peridynamics are briefly reviewed in section \ref{S2}. In section \ref{S3}, the integral domain for the peridynamics is reconstructed, and a new type of element, called peridynamic element (PE), is defined in this domain, then the PEs and CEs are used to discretize the total potential energy spatially. Then, a linear algebraic equation about the total nodal displacement is obtained. In section \ref{S4}, the numerical algorithm of the peridynamics-based finite element method is described. In section \ref{S5}, numerical examples are tested to verify our formulations. Some conclusions are drawn in section \ref{S6}.

%---------------------2---------------------%
\section{Bond-based peridynamics}\label{S2}
\subsection{Basic formulation}
The bond-based peridynamic model is proposed by Silling in \cite{1silling2000reformulation}, which assumes that a point $\b{x}$ interacts with all the points
in its neighborhood, $H_{\delta(\b{x})}$, where $\delta$ is a horizon that denotes the cut-off radius of the interaction. Based on this, the equation of motion at point $\b{x}$ yields
\begin{equation}\label{motion}
	\int_{H_{\delta(\b{x})}} \b{f}(\b{\xi}) \r{d}V_{\b{\xi}}+\b{b}(\b{x})=\rho\ddot{\b{u}}\left(\b{x}\right),
\end{equation}
where $\b{b}(\b{x})$ is the external body force field, $\b{\xi} = \b{x'}-\b{x}$ is a relative position vector referred to as a bond, and $ \b{f}(\b{\xi})$ is the pairwise force field of the bond $\b{\xi}$. Particularly, $\b{f}(\b{\xi})$ denotes the nonlocal force vector that point $\b{x}'$ exerts on point $\b{x}$. For quasi-static problems, the equilibrium equation at point $\b{x}$ can be obtained by setting $\ddot{\b{u}}\left(\b{x}\right)=0$ in Eq. \eqref{motion}.

To ensure the balance of the linear momentum, the pairwise force vector $\b{f}(\b{\xi})$ must be anti-symmetric \cite{1silling2000reformulation}, i.e., $\b{f}(\b{\xi}) = -\b{f}(-\b{\xi}')$. Thus, the pairwise force is assumed to be central \cite{2lubineau2012morphing}, i.e.
\begin{equation}\label{key}
	\b{f}(\b{\xi}) = \hat{\b{f}}[\b{x}]\left\langle \b{\xi} \right\rangle - \hat{\b{f}}[\b{x}']\left\langle -\b{\xi} \right\rangle,
\end{equation}
where $\hat{\b{f}}[\b{x}]\left\langle \b{\xi} \right\rangle$ (respectively $\hat{\b{f}}[\b{x}']\left\langle -\b{\xi} \right\rangle$) is the partial interaction due to the action of
point $\b{x'}$ over point $\b{x}$ (with respect to point $\b{x}$ over point $\b{x'}$). For homogeneous materials with linear elasticity and small deformations, a possible constitutive model \cite{1silling2000reformulation,3silling2010peridynamic} is
\begin{equation}\label{key}
	\hat{\b{f}}[\b{x}]\left\langle \b{\xi} \right\rangle = \frac{1}{2}\b{C}(\b{\xi}) \cdot \left(\b{u}(\b{x'}) - \b{u}(\b{x})\right),
\end{equation}
where $\b{u}$ is the displacement field. $\b{C}(\b{\xi})$ is the micromodulus tensor of the bond $\b{\xi}$, which is defined as \cite{1silling2000reformulation}
\begin{equation}\label{key}
\b{C}(\b{\xi})=c(\|\b{\xi}\|) \frac{\b{\xi} \otimes \b{\xi}}{\|\b{\xi}\|^2},	
\end{equation}
where $c(\|\b{\xi}\|)$ is the micromodulus coefficient.

To sum up, for quasi-static problems, the governing equations of the bond-based peridynamics for a configuration $\Omega$, $\Omega \subset \mathbb{R}^d(d = 1, 2, 3)$, can be summarized as
\begin{subequations}\label{gov}
	\begin{numcases}{}
		\int_{H_{\delta(\b{x})}} {\b{f}\left(\b{\xi}\right)} \r{d}V_{\b{\xi}}+\b{b}(\b{x})=\mathbf{0},\quad \forall \b{x} \in \Omega, \label{ga}\\
		\b{f}\left(\b{\xi}\right)  = \b{C}(\b{\xi}) \cdot \b{\eta}(\b{\xi}),\quad \forall \b{x'} \in H_{\delta(\b{x})}, \b{x} \in \Omega, \label{gb}\\
		\b{\eta}(\b{\xi}) = \b{u}(\b{x'}) - \b{u}(\b{x}),\quad \forall \b{x'}, \b{x} \in \Omega \quad\text{ and }\quad \b{u}(\b{x}) = {\b{u}}^*(\b{x}), \quad \forall \b{x} \in \partial\Omega_{\b{u}}.\label{gc}
	\end{numcases}
\end{subequations}
where ${\b{u}}^*$ is the prescribed displacement on $\partial\Omega_{\b{u}}$, $\b{\eta}$ is a measure of deformation of bond $\b{\xi}.$ Eq. \eqref{ga}, Eq. \eqref{gb} and Eq. \eqref{gc} are the \emph{static admissibility}, \emph{constitutive equation} and \emph{kinematic admissibility and compatibility}, respectively.

\subsection{Failure of the bond and structure}
In the peridynamics, once the failure of the structure is considered, the bond break is needed. Different kinds of bond break criteria have been introduced in the literature \cite{6silling2005meshfree,7silling2010crack,8foster2011energy}, and here we adopt the stretch-based criterion proposed by Silling and Askari \cite{6silling2005meshfree}. The stretch of a bond $\b{\xi}$ is defined as
\begin{equation}\label{key}
	s = \frac{\left\| \b{\xi} + \b{\eta}\right\| - \left\| \b{\xi} \right\|}{\left\| \b{\xi} \right\|}.
\end{equation}
Then the bond break can be recorded with a history-dependent scalar-valued function, $\mu$, which is defined as
\begin{equation}\label{crit}
	\mu(\b{\xi}, t)=\left\{\begin{array}{ll}
		1, & \text { if } s\left(\b{\xi}, \tau\right)<s_{c r i t}, \text { for all } 0 \le \tau \le t, \\
		0, & \text { otherwise },
	\end{array}\right.
\end{equation}
where $t$ and $\tau$ denote computational steps. $s_{c r i t}$ is the critical bond stretch. Note that Eq. \eqref{crit} means the brittle fracture mode. Multiply $\mu(\b{\xi}, t)$ to the right-hand side of Eq. \eqref{gb}, and then the constitutive equation including bond break is obtained.

Based on $s_{c r i t}$, the critical energy dissipation of the broken bond yields \cite{9wang2021strength}
\begin{equation}\label{key}
	\omega_{c r i t}=\frac{1}{2} c(\|\b{\xi}\|) s_{c r i t}^{2}\|\b{\xi}\|^{4}.
\end{equation}
Note that the critical energy dissipation $\omega_{c r i t}$ is different for bonds with different lengths. With $\mu$ and $\omega_{c r i t}$ at hand, the effective damage for each point $\b{x}$ is defined as \cite{9wang2021strength}
\begin{equation}\label{key}
	\phi(\b{x})=\frac{\int_{H_{\delta}(\b{x})}(1-\mu(\b{\xi}, t)) \omega_{c r i t} \r{d} V_{\xi}}{\int_{H_{\delta}(\b{x})} \omega_{c r i t} \r{d} V_{\xi}},
\end{equation}
which then indicates the failure of the structure.

\subsection{Principle of minimum potential energy}
In 2016, Han et al.\cite{4han2016morphing} derived the principle of minimum potential energy for a hybrid classical continuum mechanics and state-based peridynamic model. As a special case of the hybrid model, the principle of minimum potential energy for the bond-based peridynamic model can be directly obtained as: the solution of Eq. \eqref{gov} is also the solution of 
\begin{equation}\label{key}
	\arg \left\lbrace \min\limits_{\b{u} \in \mathcal{U}(\Omega)} \Pi(\b{u})\right\rbrace,
\end{equation}
where
\begin{equation}\label{key}
	\mathcal{U}(\Omega) := \left\lbrace \b{u} \in L^2(\Omega): \b{u} = \b{{u}}^* \text{ on } \partial\Omega_{\b{u}}\right\rbrace 
\end{equation}
is the kinematically admissible displacement space \cite{5azdoud2014morphing}, and the total potential energy $\Pi(\b{u})$ is defined as
\begin{equation}\label{ene}
	\Pi(\b{u}) = \frac{1}{4}\int_{\Omega}{\int_{H_{\delta(\b{x})}}{\b{f}(\b{\xi}) \cdot \b{\eta}(\b{\xi}) \r{d}V_{\b{\xi}}}  \r{d}V_{\b{x}}} - \int_{\Omega}{\b{u}(\b{x}) \cdot \b{b}(\b{x}) \r{d}V_{\b{x}}},
\end{equation}
where the first and second terms on the right-hand side of Eq. \eqref{ene} are the deformation energy and external work, respectively.

Besides, it has been proved in \cite{4han2016morphing} that the sufficient and necessary condition for $\Pi(\b{u})$ to reach the minimum yields $\delta\Pi(\b{u}) = \b{0}$, i.e., 
\begin{equation}\label{min-con}
	\frac{\partial \Pi(\b{u})}{\partial \b{u}}\delta\b{u} = \b{0}, \quad \forall \delta\b{u} \in \mathcal{V}(\Omega),
\end{equation}
where
\begin{equation}\label{key}
	\mathcal{V}(\Omega) := \left\lbrace \b{u} \in L^2(\Omega): \b{u} = \b{0} \text{ on } \partial\Omega_{\b{u}}\right\rbrace 
\end{equation}
is the trial space.

%---------------------3---------------------%
\section{Peridynamics-based finite element method}\label{S3}
\subsection{Reconstruction of the formulation for potential energy}
The classical FEM is based on the classical continuum mechanics, in which the expression of the total potential energy is a single integral on the configuration $\Omega$. However, in the peridynamics, the expression of the total potential energy contains a double integral term, i.e., the deformation energy of the structure, see Eq. \eqref{ene}, which results in the inconsistency of the finite element framework for the peridynamics with that of the classical FEM.

To avoid the above issue, we reconstruct the formulation of the deformation energy. Firstly, note that $\b{f}(\b{\xi}) = 0, \forall \b{x'} \notin H_{\delta(\b{x})}$, thus the inner integral defined on $H_{\delta(\b{x})}$ can be extended to the entire configuration $\Omega$. Therefore, the deformation energy of $\Omega$ can be rewritten as
\begin{equation}\label{key}
	\Pi_1(\b{u}) = \frac{1}{4}\int_{\Omega}{\int_{\Omega}{\b{f}(\b{\xi}) \cdot \b{\eta}(\b{\xi }) \r{d}V_{\b{\xi}}}  \r{d}V_{\b{x}}}.
\end{equation}
Further, we define a new type of integral operation as
\begin{equation}\label{key}
	\int_{\bar{\Omega}}{\bar{\b{g}}(\b{x'}, \b{x}) \r{d}\bar{V}_{\b{x'}\b{x}}} := \int_{\Omega}{\int_{\Omega}{\b{g}(\b{\xi}) \r{d}V_{\b{\xi}}} \r{d}V_{\b{x}}},
\end{equation}
where $\bar{\Omega}$ is an integral domain generated by two $\Omega$s, $\bar{\b{g}}(\b{x'},\b{x})$ is a double-parameter function related to $\b{g}(\b{\xi})$ and defined on $\bar{\Omega}$. Then the total potential energy $\Pi(\b{u})$ can be presented in a single integral form, i.e. 
\begin{equation}\label{single-ene}
	\Pi(\b{u}) = \frac{1}{4}\int_{\bar{\Omega}}{\bar{\b{f}}(\b{x'}, \b{x}) \cdot \bar{\b{\eta}}(\b{x'}, \b{x}) \r{d}\bar{V}_{\b{x'}\b{x}}} - \int_{\Omega}{\b{u}(\b{x}) \cdot \b{b}(\b{x}) \r{d}V_{\b{x}}}.
\end{equation}

\subsection{Peridynamic element}
In this section, we introduce the peridynamic elements (PEs) in the domain $\bar{\Omega}$ to preserve the single integral characteristic when discretizing the total potential energy. 

Before introducing PEs, we briefly review the elements in classical FEM because the PEs are generated based on them. In classical FEM, the configuration $\Omega$ is divided by a set of elements, $\left\lbrace e_i \right\rbrace_{i=1}^{m}$, where $m$ is the total number of the elements. These elements are not overlapped but share edges and vertices, called nodes, between adjacent elements. Therefore, we name the element in classical FEM as continuous element (CE) to distinguish it from peridynamic element (PE). The set of mesh nodes is denoted as $\left\lbrace \r{P}_l \right\rbrace_{l=1}^{n}$, with $n$ the total number of the nodes.

\begin{figure*}[htbp] 
	\centering 
	\begin{minipage}[b]{0.9\linewidth} 
		\subfloat[]{
			\begin{minipage}[b]{0.33\linewidth}
				\centering
				\includegraphics[width=0.85\linewidth]{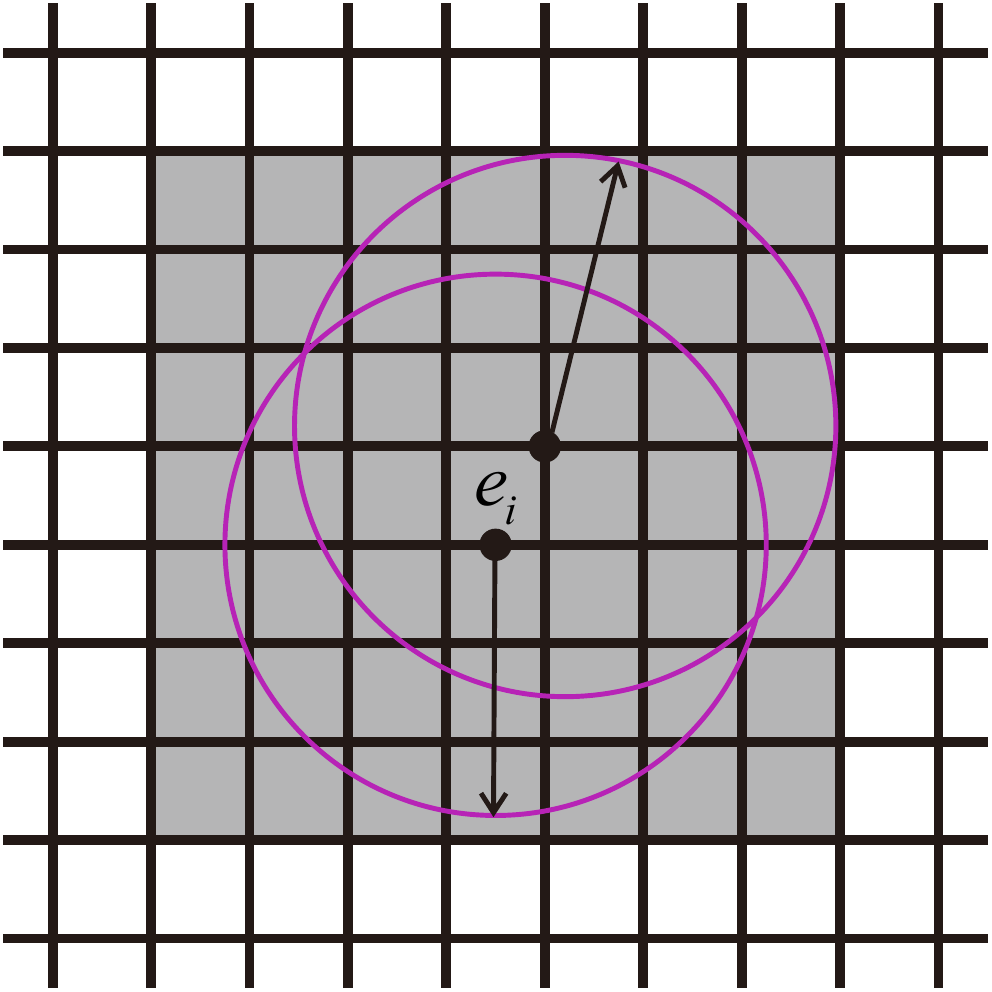}
			\end{minipage}
			\label{fig01a}
		}
		%\hfill
		\subfloat[]{
			\begin{minipage}[b]{0.33\linewidth}
				\centering
				\includegraphics[width=0.85\linewidth]{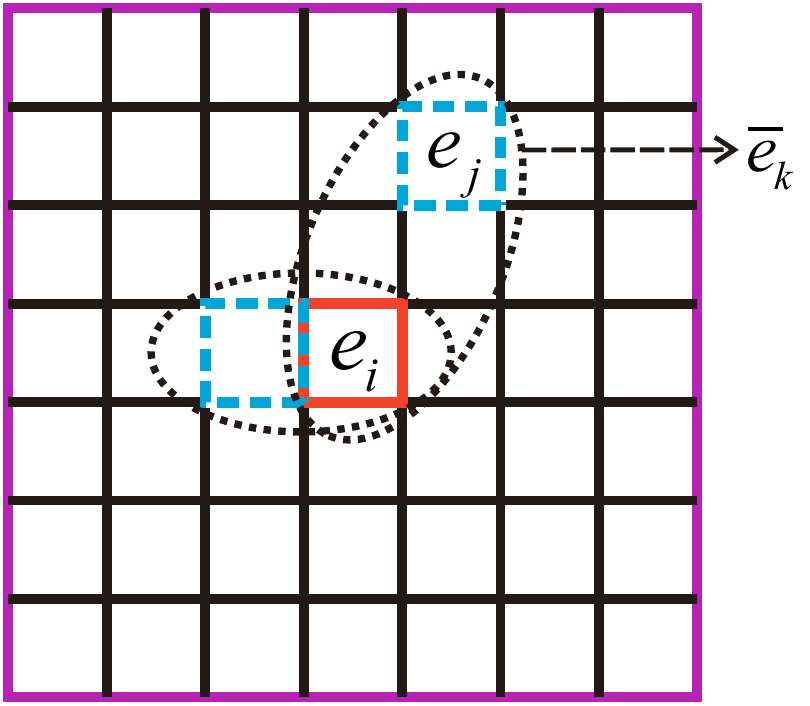}
			\end{minipage}
			\label{fig01b}
		}
		%\hfill
		\subfloat[]{
			\begin{minipage}[b]{0.33\linewidth}
				\centering
				\includegraphics[width=0.85\linewidth]{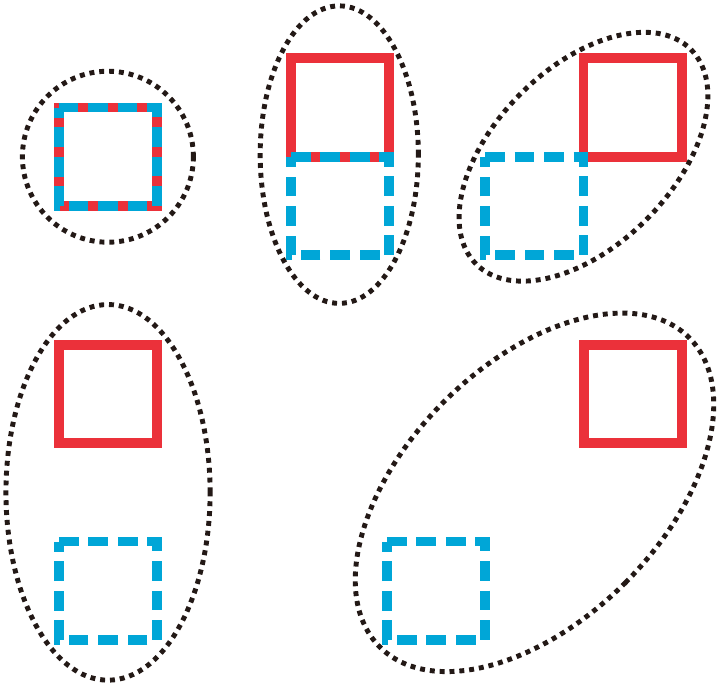}
			\end{minipage}
			\label{fig01c}
		}
		\vfill
		\caption{Neighborhood $H_i$ of CE $e_i$ and the PEs in $H_i$ for the 2D case. (a) shows the neighborhood $H_i$, the gray region, of CE $e_i$; (b) displays some PEs generated by $e_i$ and $e_j \in H_i$; (c) shows the PEs with different configurations.}
		\label{fig01}
	\end{minipage}
\end{figure*}

Fig. \ref{fig01} shows the relation between PEs and CEs in the 2D case. Suppose $\Omega$ is divided into a set of CEs, $\left\lbrace e_i \right\rbrace_{i=1}^{m}$. For each $e_i$, its neighborhood $H_i$ is defined as the minimal coverage of the union of the neighborhood of all points in $e_i$, as shown in Fig. \ref{fig01a}. Then for each CE $e_j \in H_i$, including $e_i$, there will be a new PE, denoted as $\bar{e}_k$, combined by  $e_j$ and $e_i$, as shown in Fig. \ref{fig01b}. Fig. \ref{fig01c} displays some PEs in detail, one can find that PEs may have different geometries, but they are all composed of two CEs. 

In general, if we assume that CE $e_i$ has $n_i$ nodes and the nodes labels are $\left[ {\begin{array}{*{20}{c}}
		{{\r{P}_{{i_1}}}}&{{\r{P}_{{i_2}}}}& \cdots &{{\r{P}_{{i_{{n_i}}}}}}
\end{array}} \right]$, CE $e_j$ has $n_j$ nodes and the nodes labels are $\left[ {\begin{array}{*{20}{c}}
{{\r{P}_{{j_1}}}}&{{\r{P}_{{j_2}}}}& \cdots &{{\r{P}_{{j_{{n_j}}}}}}
\end{array}} \right]$, where $\r{P}_{{\alpha}_{\beta}} \in \left\lbrace \r{P}_l \right\rbrace_{l=1}^{n} (\alpha=i,j; \beta=1, 2, \cdots, n_{\alpha})$, then PE $\bar{e}_k$ is an element with $\bar{n}_k = n_i + n_j$ nodes and the nodes labels are $\left[ {\begin{array}{*{20}{c}}{{\r{P}_{{k_1}}}}&{{\r{P}_{{k_2}}}}& \cdots &{{\r{P}_{{k_{{\bar{n}_k}}}}}}\end{array}} \right] = \left[ {\begin{array}{*{20}{c}}
{{\r{P}_{{j_1}}}}&{{\r{P}_{{j_2}}}}& \cdots &{{\r{P}_{{j_{{n_j}}}}}}&{{\r{P}_{{i_1}}}}&{{\r{P}_{{i_2}}}}& \cdots &{{\r{P}_{{i_{{n_i}}}}}}
\end{array}} \right]$. Based on the CEs, $\left\lbrace e_i \right\rbrace_{i=1}^{m}$, and the above method of generating PE, a set of PEs, $\left\lbrace \bar{e}_k \right\rbrace_{k=1}^{\bar{m}}$, will finally be obtained. For example, $n_i = n_j = 4$ and $\bar{n}_k = 8$ in Fig. \ref{fig01c}. 

\begin{remark}
	 Each PE is composed of two and only two CEs, no matter for the 1D, 2D or 3D cases, which is consistent with $\bar{\Omega}$ consisting of two $\Omega$s. Only in this way could the discrete format of Eq. \eqref{single-ene} maintains the characteristics of the single integral form.
\end{remark}
\begin{remark}
	Even $e_i$ and $e_j$ that make up $\bar{e}_k$ have same nodes, the duplicate nodes will not be eliminated in $\bar{e}_k$.
\end{remark}

\subsection{Spatial discretization of the potential energy based on PEs and CEs}
Now we have two sets of elements, CEs and PEs. The former is used to calculate local quantities, and the latter is used to calculate nonlocal quantities.

According to the interpolation technique of the classical FEM, the displacement field $\b{u}_i(\b{x})$ on any CE $e_i$ can be approximately expressed as
\begin{equation}\label{dis-FE}
	\b{u}_i(\b{x}) = \b{N}_i(\b{x})\b{d}_i,
\end{equation}
where
\begin{small}
\begin{align}\label{key}
	\b{N}_i(\b{x}) &=\left[\begin{array}{cccccccccc}
		N_{i_{1}}(\b{x}) & 0 & 0 & N_{i_{2}}(\b{x}) & 0 & 0 & \cdots & N_{i_{n_{i}}}(\b{x}) & 0 & 0 \\
		0 & N_{i_{1}}(\b{x}) & 0 & 0 & N_{i_{2}}(\b{x}) & 0 & \cdots & 0 & N_{i_{n_{i}}}(\b{x}) & 0 \\
		0 & 0 & N_{i_{1}}(\b{x}) & 0 & 0 & N_{i_{2}}(\b{x}) & \cdots & 0 & 0 & N_{i_{n_{i}}}(\b{x})
	\end{array}\right],\\
	\b{d}_i &= \left[\begin{array}{llllllllll}
		u_{i_{1}} & v_{i_{1}} & w_{i_{1}} & u_{i_{2}} & v_{i_{2}} & w_{i_{2}} & \cdots & u_{i_{n_{i}}} & v_{i_{n_{i}}} & w_{i n_{i}}
	\end{array}\right]^{T}
\end{align}
\end{small}
are the \emph{shape function matrix of CE} and the \emph{nodal displacement vector of CE} for $e_i$, respectively. Here, $\left\lbrace N_{i_l}(\b{x})\right\rbrace_{l=1}^{n_i}$ is the \emph{shape function} for node $\r{P}_{i_l}$,  $\left\lbrace u_{i_l}\right\rbrace_{l=1}^{n_i}$, $\left\lbrace v_{i_l}\right\rbrace_{l=1}^{n_i}$ and $\left\lbrace w_{i_l}\right\rbrace_{l=1}^{n_i}$ are the displacement components of node $\r{P}_{i_l}$ along the $X$, $Y$, and $Z$ directions, respectively.

The displacement field $\bar{\b{u}}_k(\b{x'},\b{x})$ on any PE $\bar{e}_k$ can be approximately expressed as
\begin{equation}\label{dis-Peri-FE}
	\bar{\boldsymbol{u}}_{k}\left(\b{x'}, \b{x}\right)=\left[\begin{array}{c}
		\b{u}_{j}\left(\b{x'}\right) \\
		\b{u}_{i}(\b{x})
	\end{array}\right]=\bar{\b{N}}_k\left(\b{x'}, \b{x}\right) \bar{\b{d}}_{k},
\end{equation}
where
\begin{align}
	\bar{\b{N}}_k(\b{x'}, \b{x}) &= \left[\begin{array}{cc}
		\boldsymbol{N}_{j}\left(\boldsymbol{x}^{\prime}\right) & \boldsymbol{0} \\
		\mathbf{0} & \boldsymbol{N}_{i}(\boldsymbol{x})
	\end{array}\right],\\
	\bar{\b{d}}_k &= \left[\begin{array}{cc}
		\b{d}_j \\
		\b{d}_i
	\end{array}\right],
\end{align}
are the \emph{shape function matrix of PE} and the \emph{nodal displacement vector of PE} for $\bar{e}_k$, respectively. 

Then, based on Eq. \eqref{gc} and Eq. \eqref{dis-Peri-FE}, for any PE $\bar{e}_k$ we have
\begin{equation}\label{diff-Peri-FE}
	\bar{\b{\eta}}_{k}\left(\b{x}^{\prime}, \b{x}\right)=\b{u}_{j}\left(\b{x'}\right)-\b{u}_{i}(\b{x})=\bar{\b{B}}_k\left(\b{x'}, \b{x}\right) \bar{\b{d}}_{k},
\end{equation}
where
\begin{equation}\label{diff-mat}
	\bar{\b{B}}_k\left(\b{x'}, \b{x}\right) = \bar{\b{H}} \bar{\b{N}}\left(\b{x'}, \b{x}\right)
\end{equation}
is the \emph{difference matrix for shape function} for $\bar{e}_k$. In Eq. \eqref{diff-mat}, 
\begin{equation}\label{key}
	\bar{\b{H}} = \left[\begin{array}{cc}
		\b{I} & -\b{I}
	\end{array}\right]
\end{equation}
is the \emph{difference operator matrix} and $\b{I}$ is an identity matrix of dimension $d\ (d=1,2,3)$. 

Next, base on Eq. \eqref{gb} and Eq. \eqref{diff-Peri-FE}, for any PE $\bar{e}_k$ we have
\begin{equation}\label{forc-Peri-FE}
	\bar{\b{f}}_{k}\left(\b{x'}, \b{x}\right)=\bar{\b{S}}_k\left(\b{x'}, \b{x}\right) \bar{\b{d}}_{k},
\end{equation}
where
\begin{equation}\label{forc-mat}
	\bar{\b{S}}_k\left(\b{x'}, \b{x}\right) = \b{D}(\b{\xi})\bar{\b{B}}_k\left(\b{x'}, \b{x}\right)
\end{equation}
is the \emph{partial interaction force matrix} for $\bar{e}_k$. In Eq. \eqref{forc-mat}, $\b{D}(\b{\xi})$ is the matrix form of the micromodulus tensor $\b{C}(\b{\xi})$. For $d = 3$ as an example, we have 
\begin{equation}\label{key}
	\b{D}(\b{\xi})=\frac{c(\left\|\b{\xi}\right\|)\mu(\b{\xi}, t)}{\left\|\b{\xi}\right\|^2} \left[\begin{array}{ccc}
		\xi_{1}^{2} & \xi_{1} \xi_{2} & \xi_{1} \xi_{3} \\
		\xi_{2} \xi_{1} & \xi_{2}^{2} & \xi_{2} \xi_{3} \\
		\xi_{3} \xi_{1} & \xi_{3} \xi_{2} & \xi_{3}^{2}
	\end{array}\right].
\end{equation}

With Eq. \eqref{dis-FE}, Eq. \eqref{dis-Peri-FE}, Eq. \eqref{diff-Peri-FE} and Eq. \eqref{forc-Peri-FE} at hand, the total potential energy $\Pi(\b{u})$ can be approximated as
\begin{equation}\label{ene-dis}
	\Pi(\b{d})=\frac{1}{4}\b{d}^T\bar{\b{K}}\b{d} - \b{d}^T\b{F},
\end{equation}
where $\b{d}$ is the \emph{total nodal displacement vector} and
\begin{align}\label{stif-mat}
	\bar{\b{K}} = \sum\limits_{k=1}^{\bar{m}}{\bar{\b{G}}_k^T\bar{\b{K}}_k\bar{\b{G}}_k },\quad \b{F} = \sum\limits_{i=1}^{m}\b{G}_i^T\b{F}_i
\end{align}
are the \emph{total stiffness matrix} and \emph{total load vector}, respectively. In Eq. \eqref{stif-mat}, $\bar{\b{G}}_k$ and $\b{G}_i$ are the \emph{transform matrix} of the degree of freedom for the nodes of $\bar{e}_k$ and the \emph{transform matrix} of the degree of freedom for the nodes of $e_i$, respectively, which satisfies
\begin{equation}\label{key}
	\bar{\b{d}}_k = \bar{\b{G}}_k\b{d}, \quad \b{d}_i = \b{G}_i\b{d},
\end{equation}
respectively. Further,
\begin{align}
	\bar{\b{K}}_k =& \int_{\bar{\Omega}_k}{\bar{\b{B}}_k^T\left(\b{x'}, \b{x}\right)\b{D}(\b{\xi})\bar{\b{B}}_k\left(\b{x'}, \b{x}\right)  \r{d}\bar{V}_{\b{x'}\b{x}}},\\
	 \b{F}_i =& \int_{\Omega_i}{\b{N}_i^T(\b{x}) \b{b}(\b{x})\r{d}V_{\b{x}}},
\end{align}
are the \emph{element stiffness matrix} and the \emph{element load vector}, respectively. 

Finally, a linear system including the solution of the nodal displacement vector $\b{d}$ can be derived from Eq. \eqref{ene-dis} using the condition Eq. \eqref{min-con}. That is
\begin{equation}\label{lineareq}
	\frac{1}{2}\bar{\b{K}}\b{d} = \b{F}.
\end{equation}

%---------------------4---------------------%
\section{Numerical algorithm}\label{S4}
\begin{figure*}[htbp]
	\centering
	\includegraphics[width=0.7\linewidth]{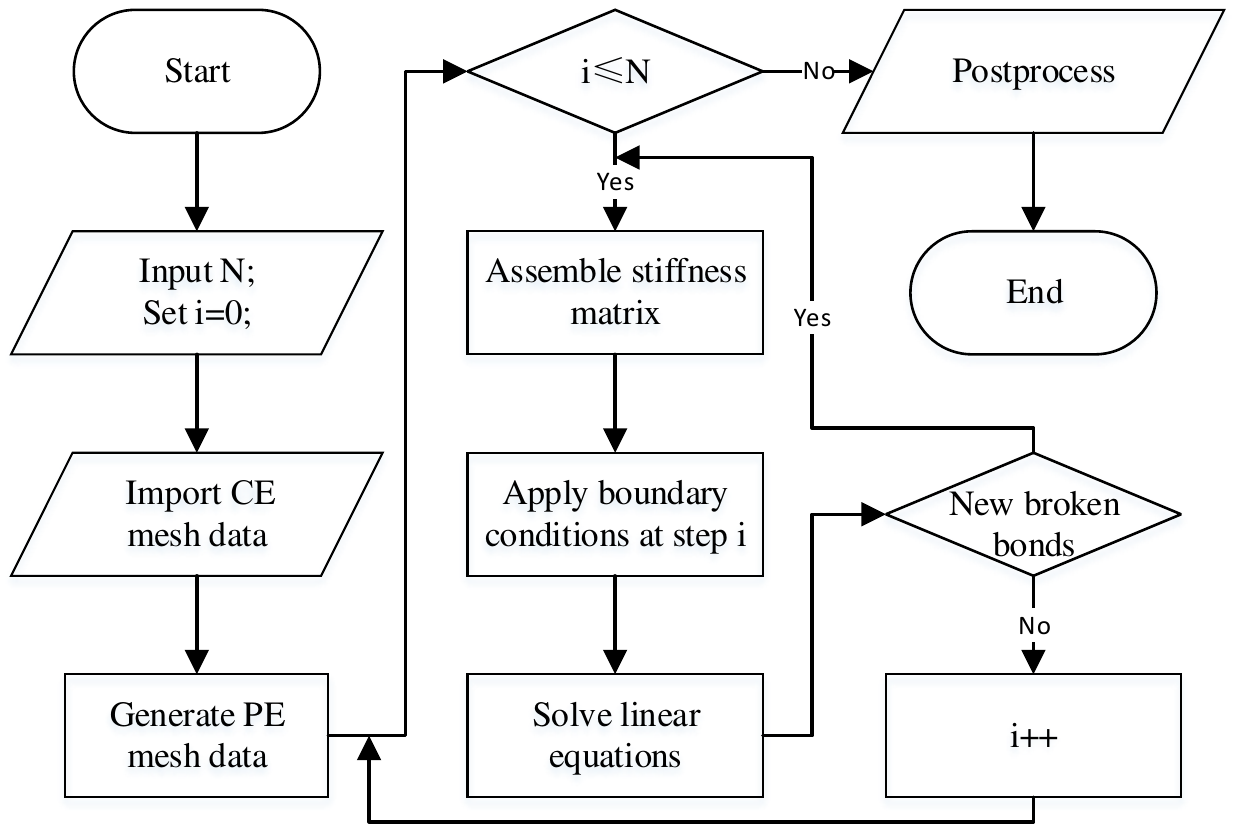}
	\caption{Flowchart of the numerical algorithm, where N is the number of total progressive increments in the simulation.}
	\label{flowchart}
\end{figure*}

This section is devoted to the numerical algorithm of the peridynamics-based finite element method for the quasi-static fracture analysis. Slightly different from the classical FEM of the proposed method is that PE mesh data must be generated after inputting the CE mesh data and before starting the fracture analysis. During the numerical simulation, the boundary conditions are specified as N progressive increments. For each incremental step, Eq. \eqref{lineareq} may be solved several times. Specifically, if new broken bonds are found after solving Eq. \eqref{lineareq}, update the stiffness matrix and solve Eq. \eqref{lineareq} again until there are no new broken bonds. For more details about the numerical algorithm, see Fig. \ref{flowchart}, the flowchart of the algorithm.

%---------------------5---------------------%
\section{Numerical examples}\label{S5}
In this section, three benchmark examples were carried out to verify the proposed peridynamics-based finite element method. In all examples, the Poisson’s ratio is fixed as $\nu=1/3$, and the horizon is chosen as $\delta = 3h$ to ensure the correct numerical integration of nonlocal effects in numerical computations, where $h$ is the average size of the CEs. The micromodulus coefficient, $c(\|\b{\xi}\|)$, is assumed to be an exponential function \cite{9wang2021strength}:
\begin{equation}\label{key}
	c(\|\xi\|)=\tau^{0} e^{-\|\xi\| / l},
\end{equation}
where $\tau^{0}$ is a constant coefficient relating to the Young's modulus and the Poisson's ratio \cite{2lubineau2012morphing, 22azdoud2013morphing}, and $l$ is a characteristic length, which is chosen as $l = \delta / 15$ in this paper. Besides, the bilinear quadrilateral CE is adopted in all examples.

\subsection{Notched beam under four-point bending}
\begin{figure*}[htbp] 
	\centering 
	\subfloat[]{
		\begin{minipage}[b]{0.9\linewidth}
			\centering
			\includegraphics[width=0.8\linewidth]{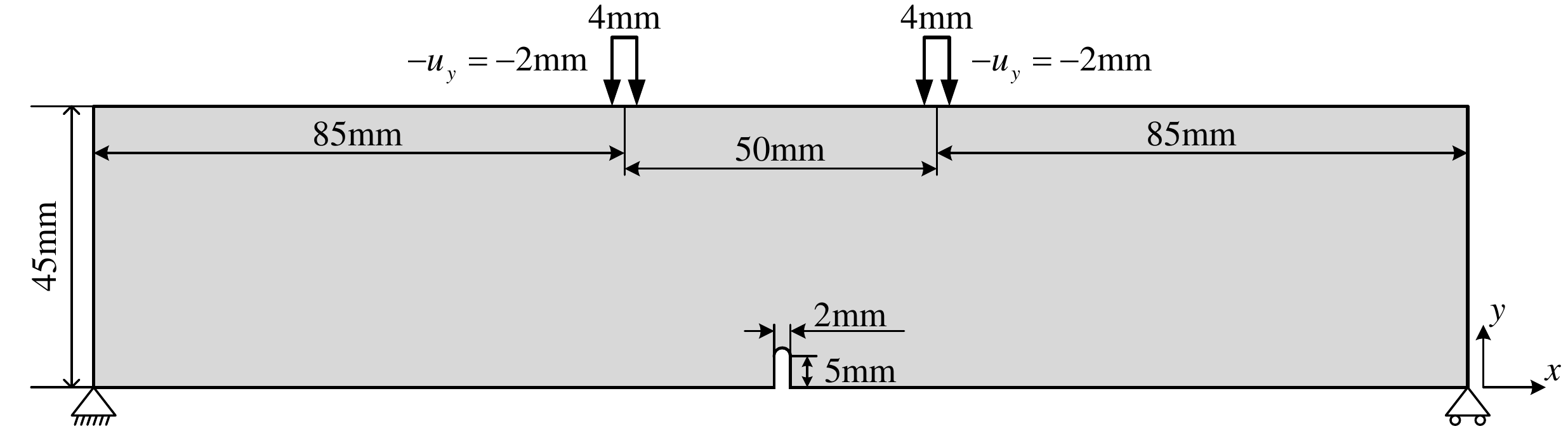}
			\label{1-1a}
		\end{minipage}
	}
	
	% 这里需要加一个回车键以达到换行的效果
	\subfloat[]{
		\begin{minipage}[b]{0.9\linewidth}
			\centering
			\includegraphics[width=0.72\linewidth]{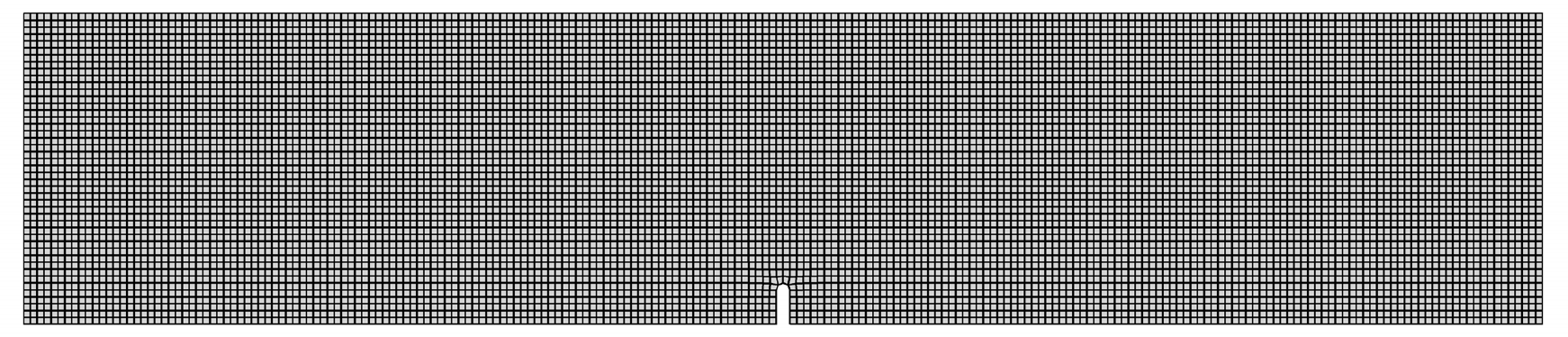}
			\label{1-1b}
		\end{minipage}
	}
	\caption{Schematics for the notched beam. (a) the geometry and loading setup; (b) the quadrilateral meshes.}
	\label{1-1}
\end{figure*}

We first conduct a four-point bending test of a single-edge notched beam. The geometry and the loading setup are shown in Fig. \ref{1-1a} and the mesh is shown in Fig. \ref{1-1b}. The Young’s modulus is $E = 30 \text{GPa}$ and the critical stretch is set to be $s_{crit} = 0.02$. The average size of the CEs is $h \approx 1.0 \text{mm}$. The simulation is implemented through 20 equably progressive increments steps, i.e., $\text{N}=20$.

\begin{figure*}[htbp] 
	\centering 
	\begin{minipage}[b]{0.71\linewidth}
		\centering
		\includegraphics[width=\linewidth]{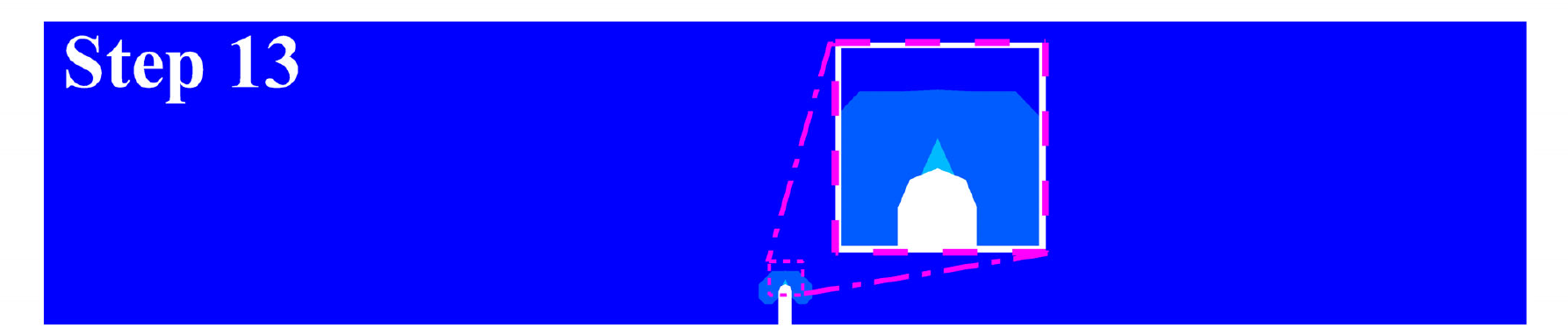}\vspace{3pt}
		\includegraphics[width=\linewidth]{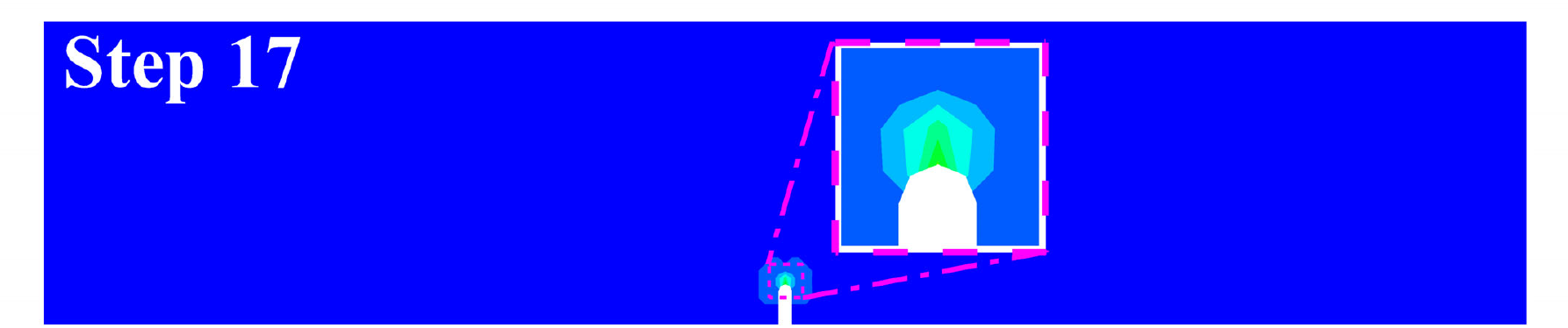}\vspace{3pt}
		\includegraphics[width=\linewidth]{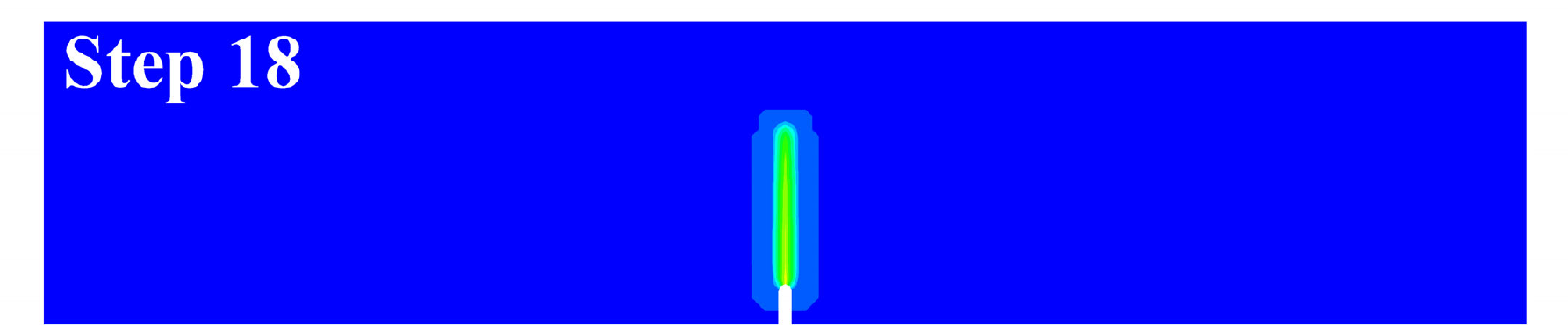}\vspace{3pt}
		\includegraphics[width=\linewidth]{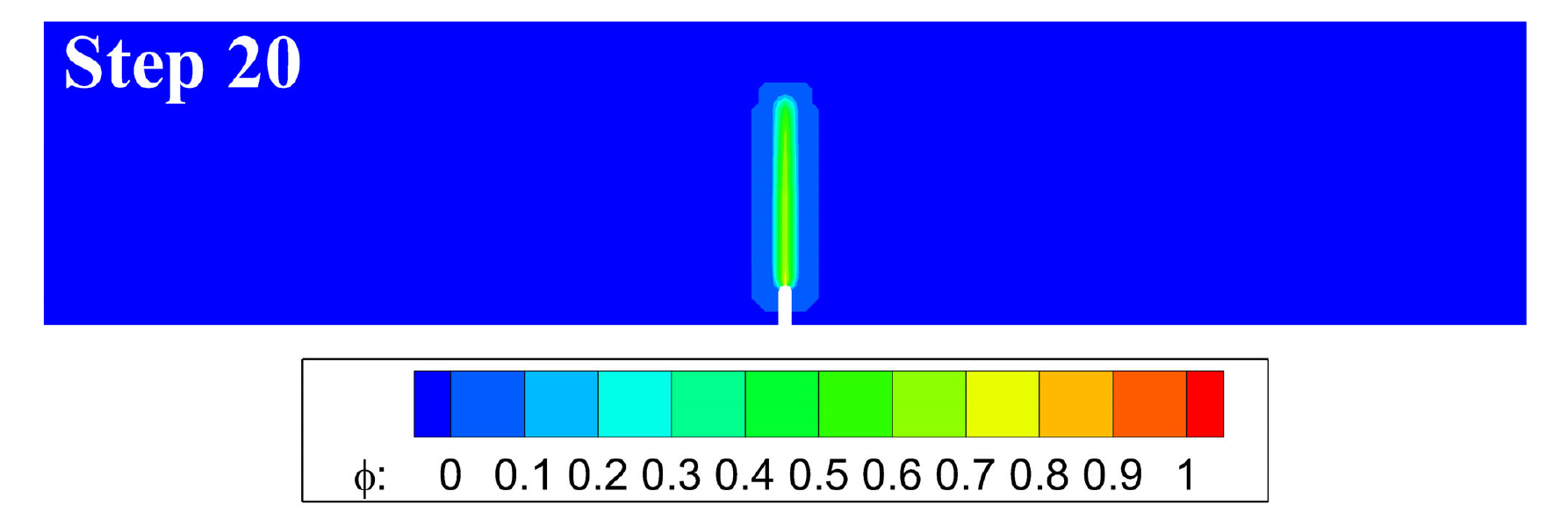}
	\end{minipage}
	\caption{Effective damage contours of the notched beam under four-point bending. The bond breaks for the first time at step 12.}
	\label{1-2}
\end{figure*}

Fig. \ref{1-2} shows the evolution of the effective damage contours during the loading process of the four-point bending for the notched beam. First, the damage initiates, i.e., the bond breaks for the first time, at step 13. Then, the damage develops slowly in the next several steps. After that, the damage propagates suddenly and violently at step 18, which is in line with brittle fracture characteristics.

\begin{figure*}[htbp]
	\centering
	\includegraphics[width=0.55\linewidth]{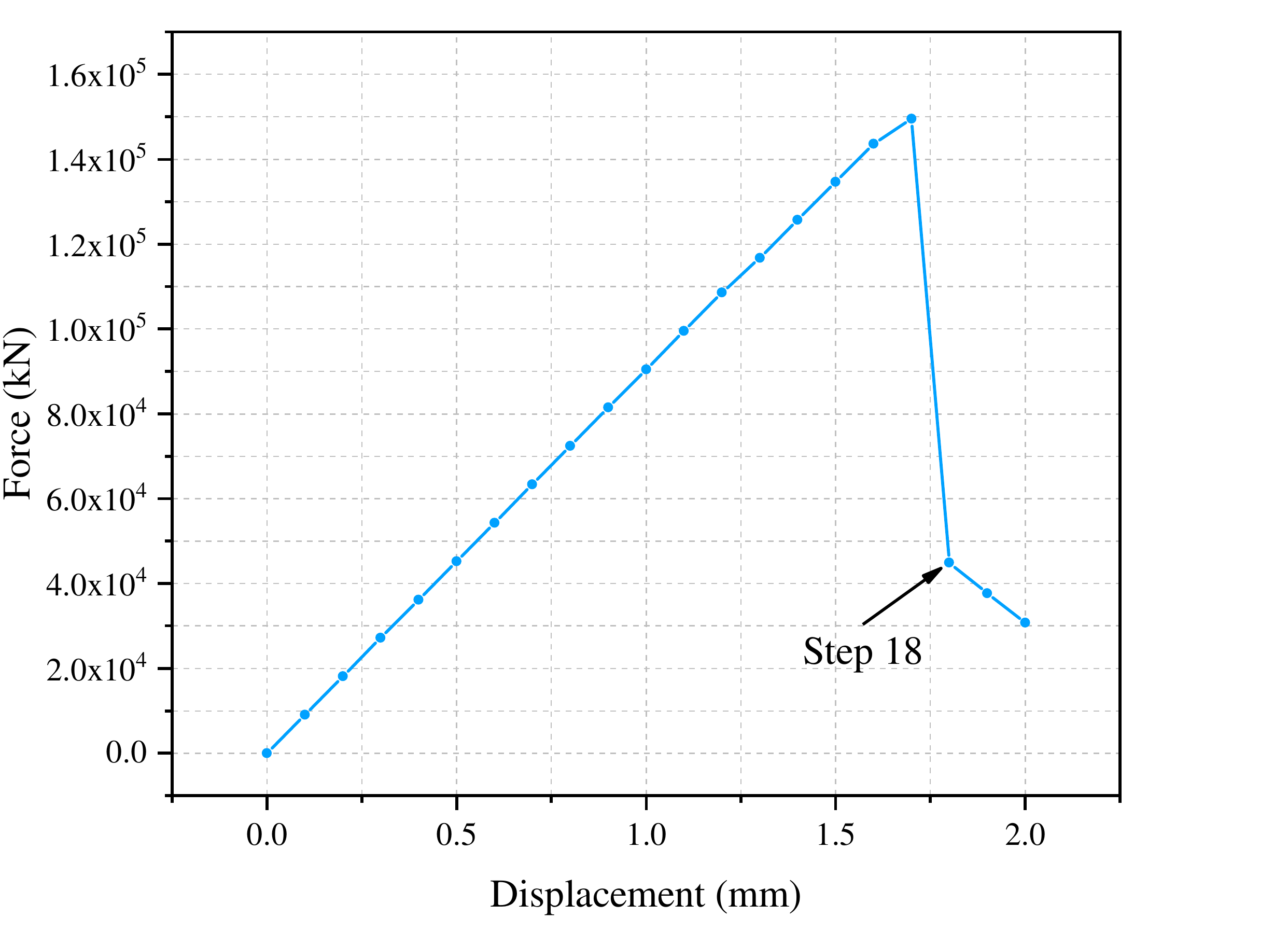}
	\caption{Force-displacement curve of the notched beam.}
	\label{1-3}
\end{figure*}

Fig. \ref{1-3} displays the curves of the average force on the loading points on the top of the beam versus the displacement. Before step 17, the effective damage is small, so  the curve approximates linear. Then between step 17 and step 18, the curve drops drastically, which is related to the sudden propagation of the crack between these two steps.

\subsection{Central notched Brazilian disk under compression}
\begin{figure*}[htbp]
	\centering
	\subfloat[]{
		\begin{minipage}[b]{0.45\linewidth}
			\centering
			\includegraphics[width=0.9\linewidth]{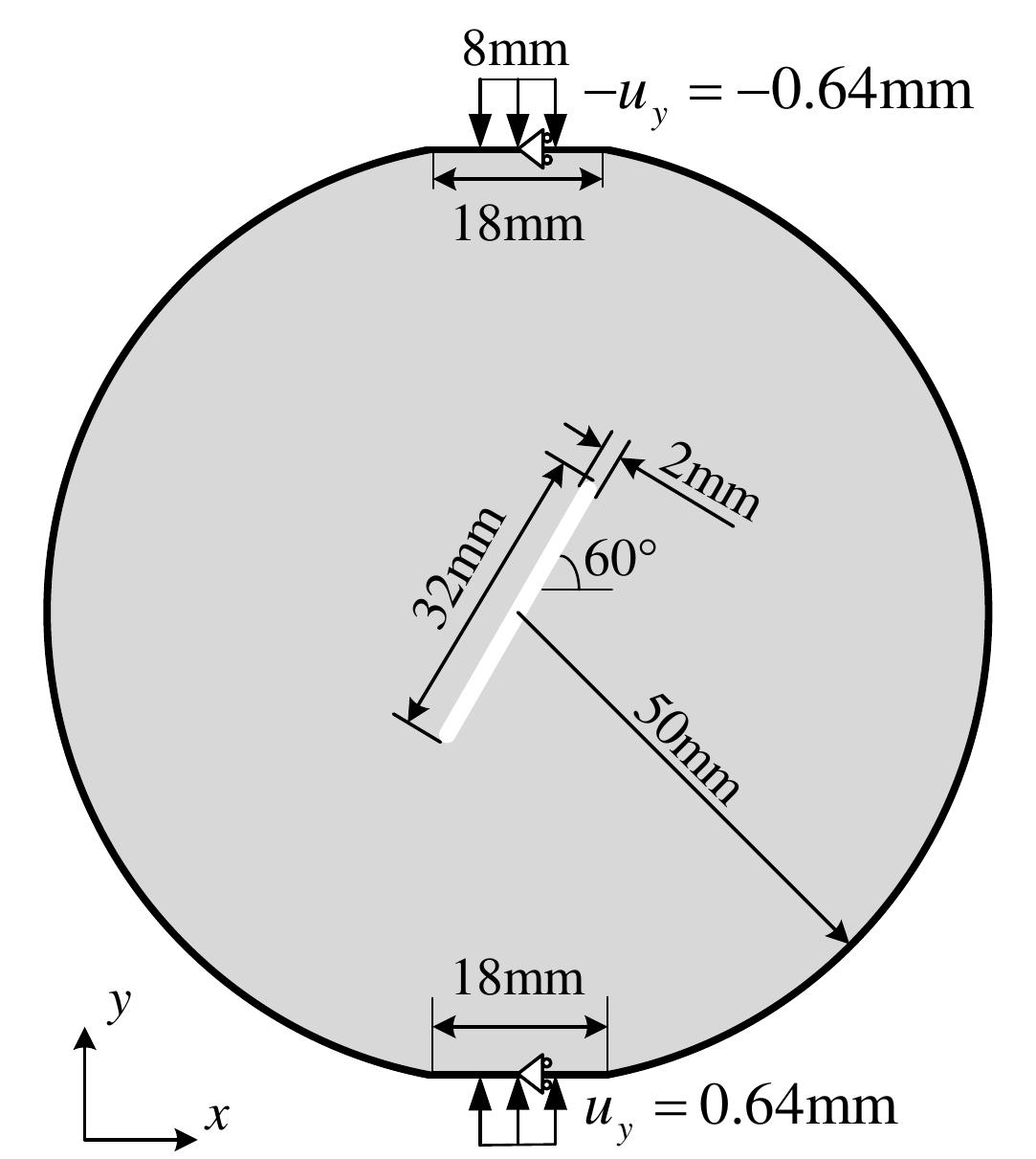}
			\label{2-1a}
		\end{minipage}
	}
	\subfloat[]{
		\begin{minipage}[b]{0.45\linewidth}
			\centering
			\includegraphics[width=0.88\linewidth]{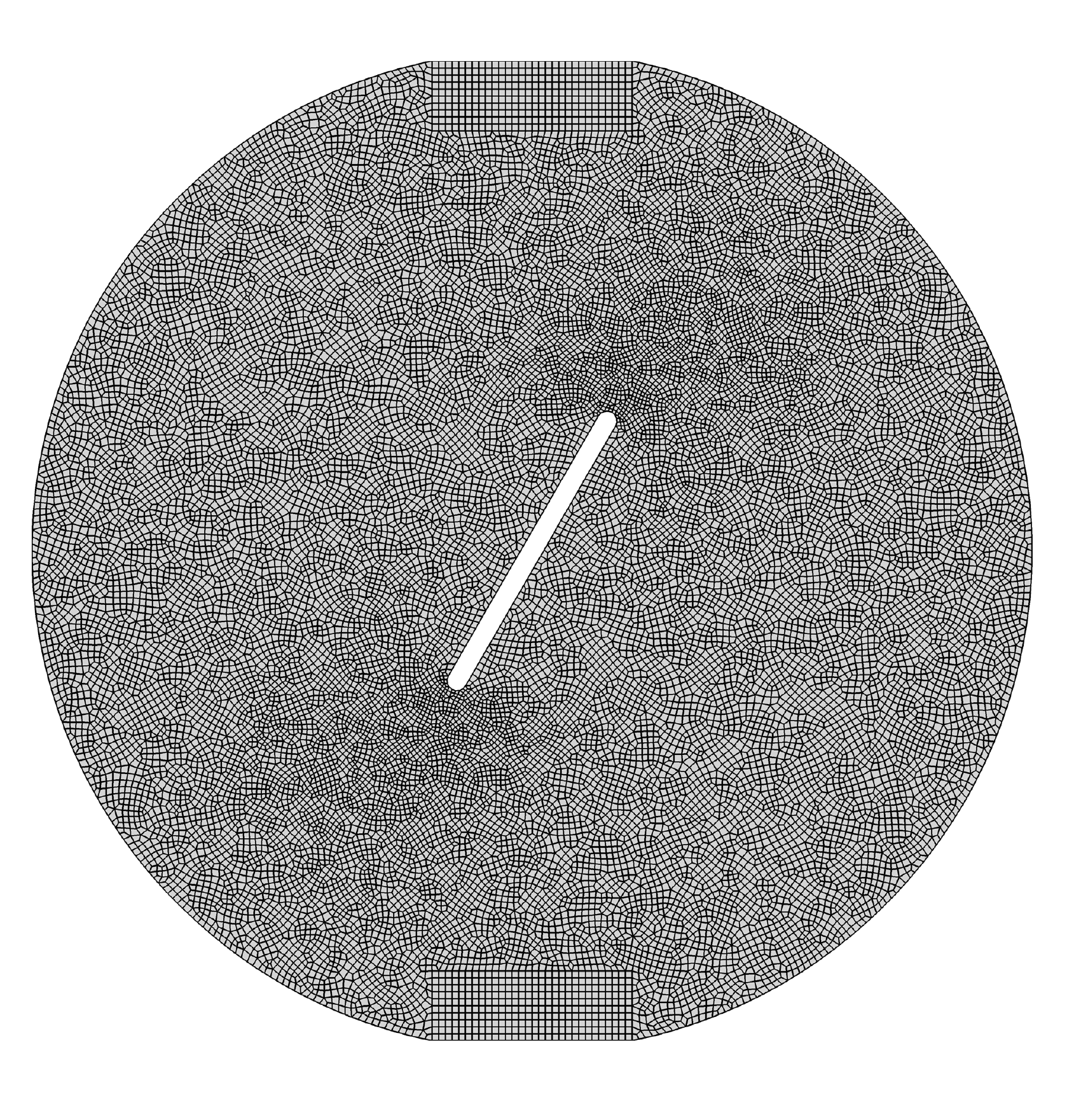}
			\label{2-1b}
		\end{minipage}
	}
	\caption{Schematics for the central notched Brazilian disk. (a) the geometry and loading setup; (b) the quadrilateral meshes.}
	\label{2-1}
\end{figure*}

We now investigate the compression test of a central notched Brazilian disk. The geometry and the loading setup of the disk are shown in Fig. \ref{2-1a} and the mesh is shown in Fig. \ref{2-1b}. The Young’s modulus is $E = 3.1 \text{GPa}$ and the critical stretch is set to be $s_{crit} = 0.02$. The average size of the CEs is $h \approx 0.7 \text{mm}$. The simulation is implemented through 64 equably progressive increments steps, i.e., $\text{N}=64$.

\begin{figure*}[htbp] 
	\centering 
	\begin{minipage}[b]{0.99\linewidth} 
		\hfill
		\subfloat{
			\begin{minipage}[b]{0.33\linewidth}
				\centering
				\includegraphics[width=\linewidth]{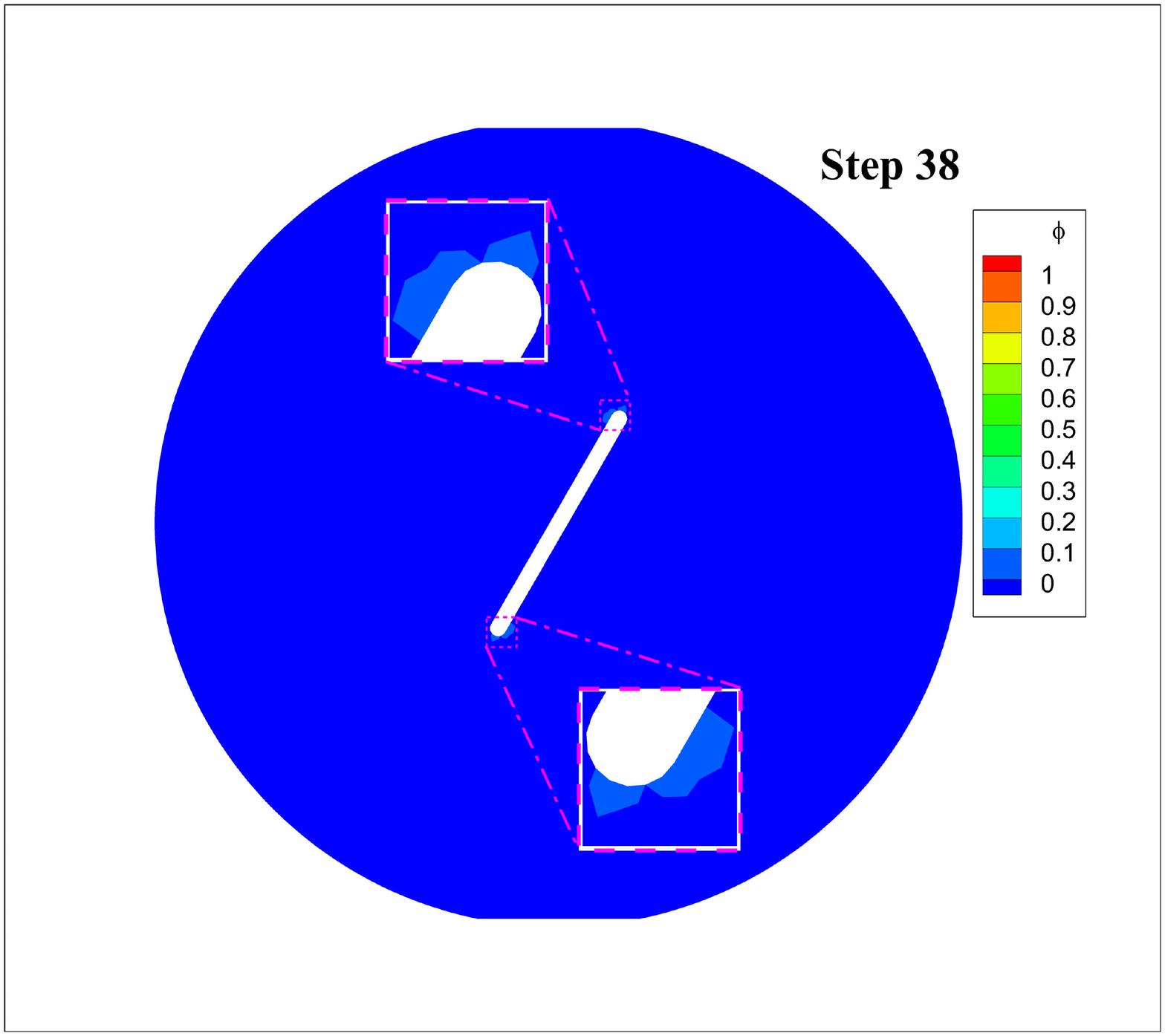}\vspace{8pt}
				\includegraphics[width=\linewidth]{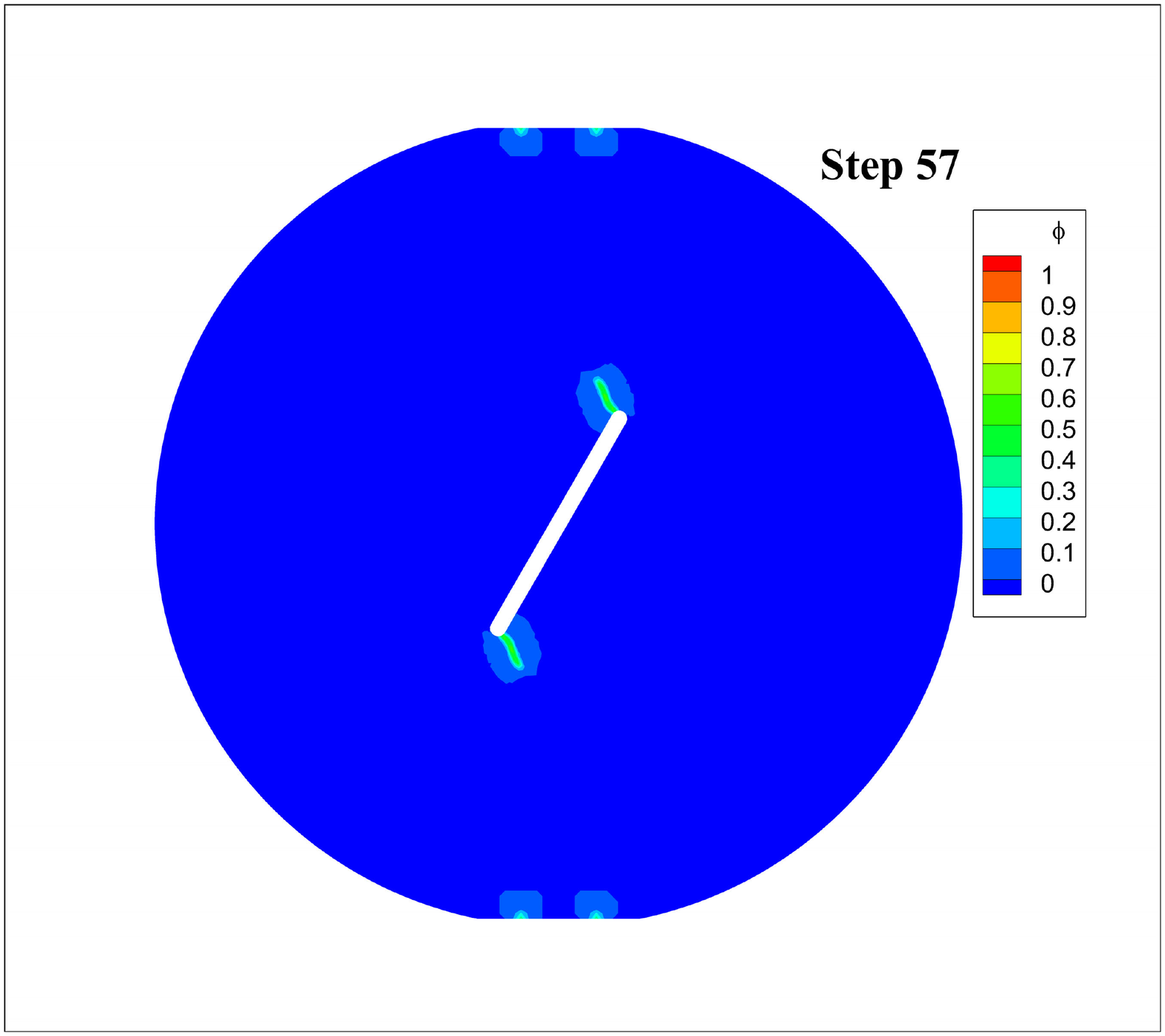}
			\end{minipage}
		}
		%\hfill
		\subfloat{
			\begin{minipage}[b]{0.33\linewidth}
				\centering
				\includegraphics[width=\linewidth]{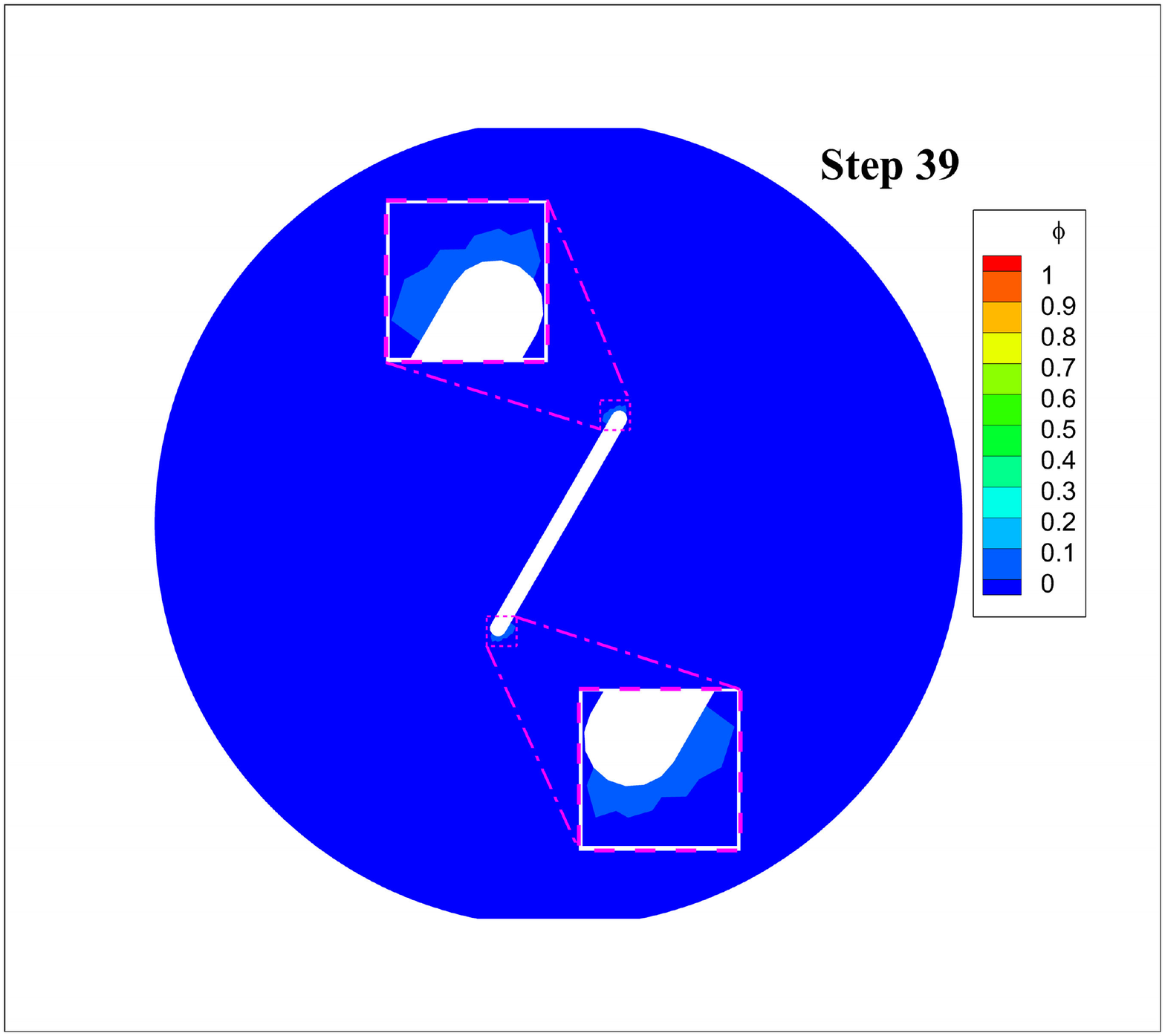}\vspace{8pt}
				\includegraphics[width=\linewidth]{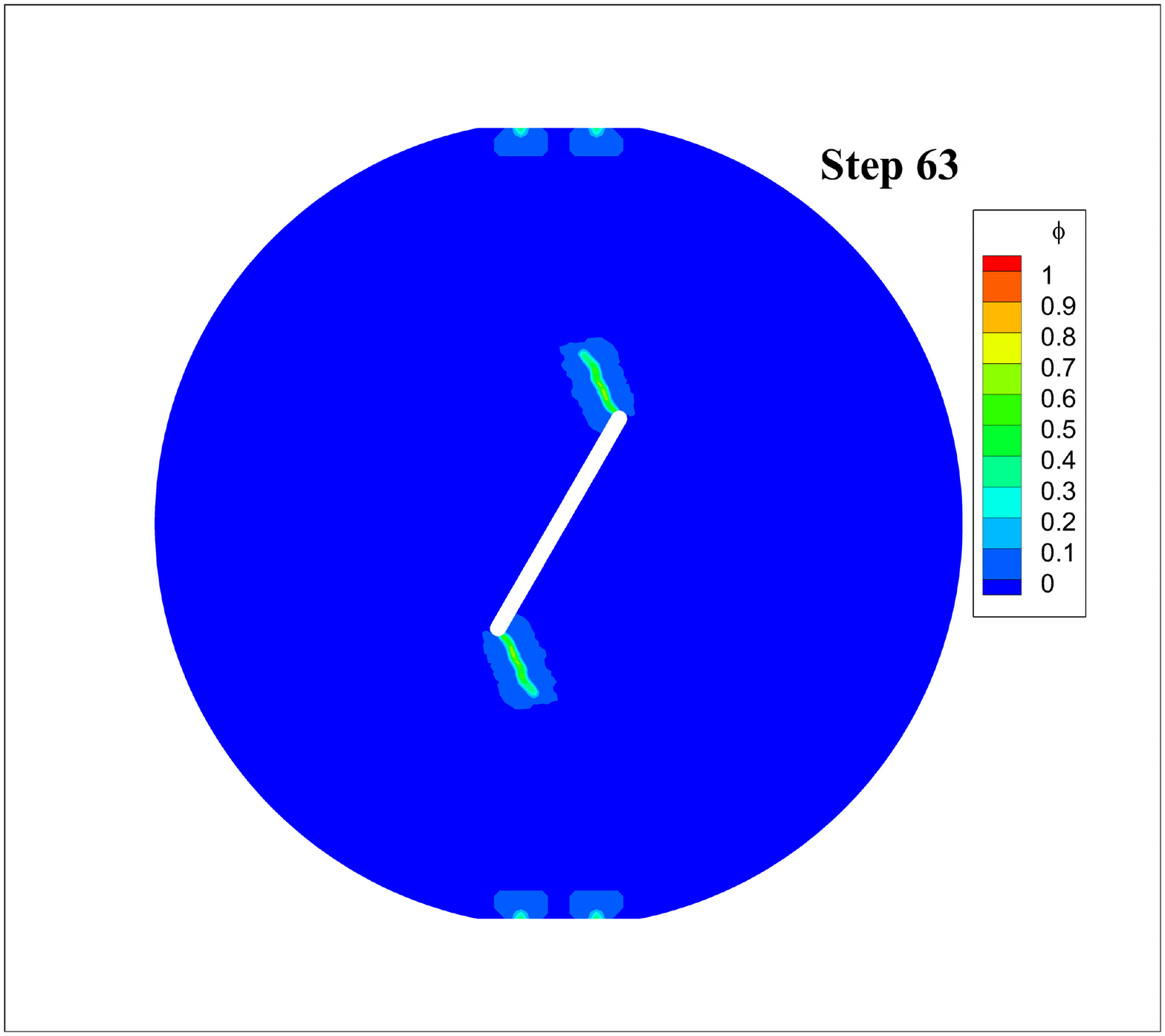}
			\end{minipage}
		}
		\subfloat{
			\begin{minipage}[b]{0.33\linewidth}
				\centering
				\includegraphics[width=\linewidth]{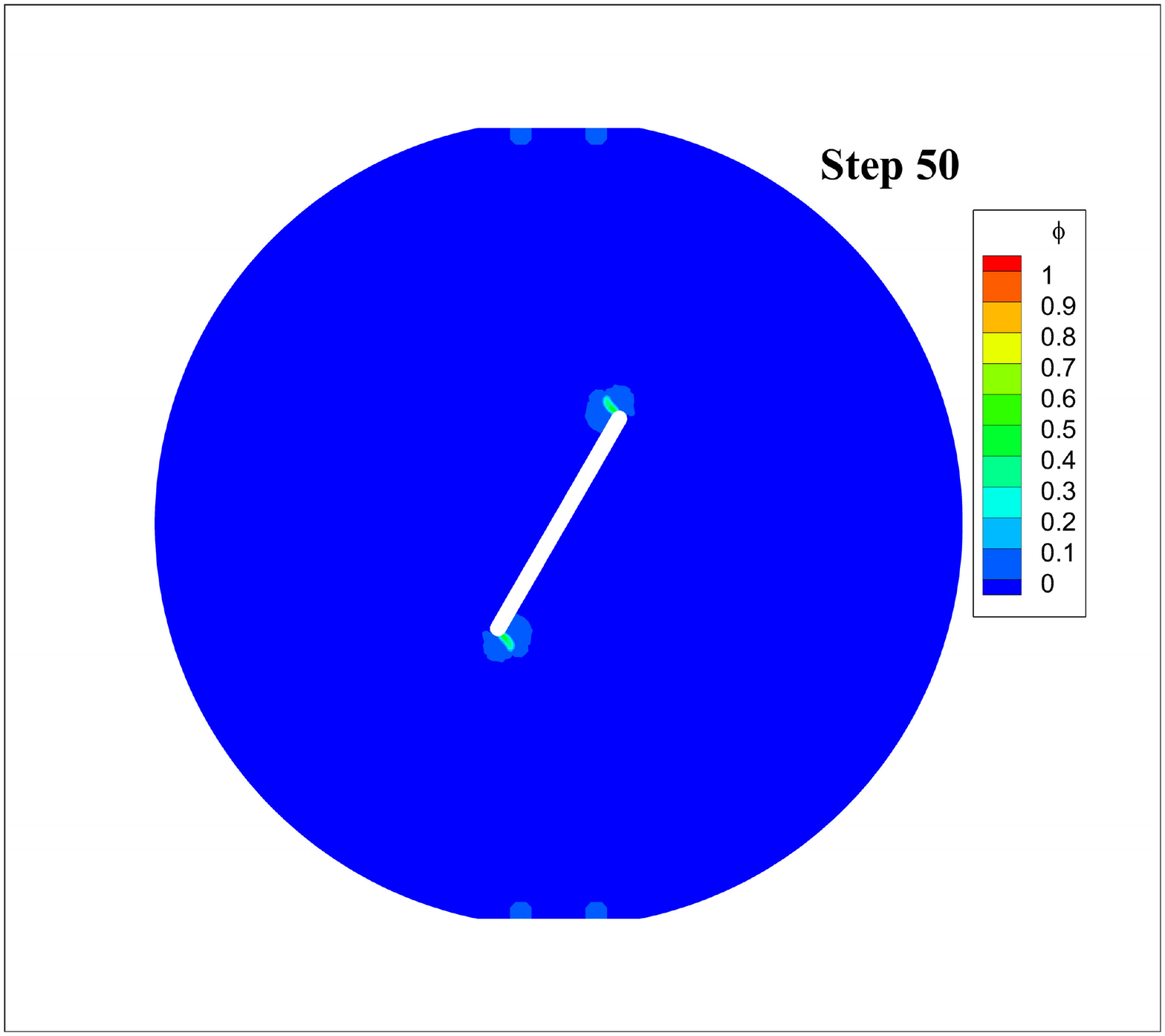}\vspace{8pt}
				\includegraphics[width=\linewidth]{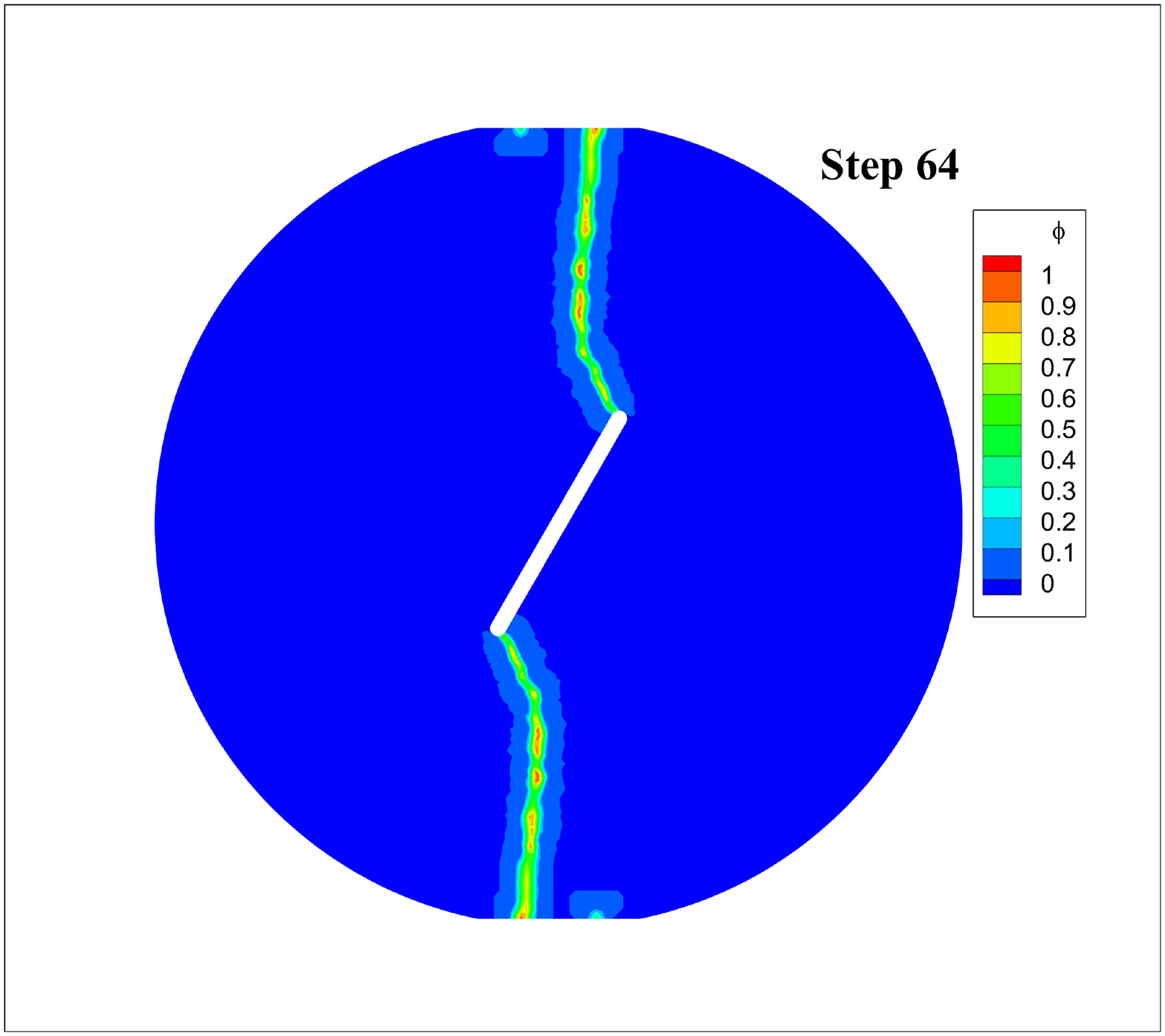}
			\end{minipage}
		}
	\end{minipage}
	\vfill
	\caption{Effective damage contours of the central notched Brazilian disk under compression.}
	\label{2-2}
\end{figure*}

Fig. \ref{2-2} displays the effective damage contours at given loading steps. The simulation results indicate that the cracks initiate at the notched corners at step 38. Then the cracks grow slowly at both corners until step 63 and suddenly penetrate the disk at step 64. The crack geometries of the disk are very similar to the experimental results reported in \cite[Fig. 17]{24haeri2014experimental}.

\begin{figure*}[htbp]
	\centering
	\subfloat[]{
		\begin{minipage}[b]{0.5\linewidth}
			\centering
			\includegraphics[height=0.7\linewidth]{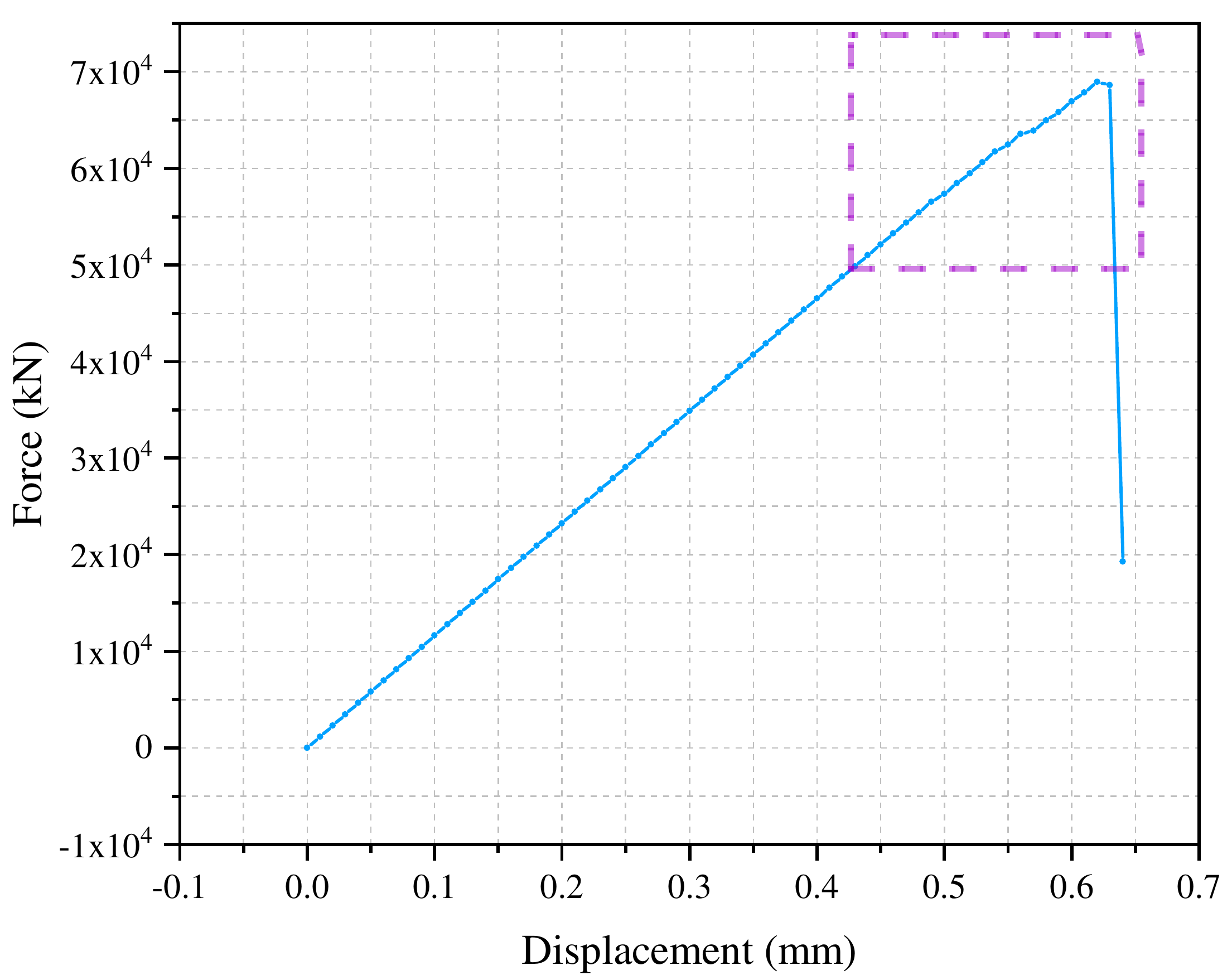}
		\end{minipage}
		\label{2-3a}
	}
	\subfloat[]{
		\begin{minipage}[b]{0.5\linewidth}
			\centering
			\includegraphics[height=0.7\linewidth]{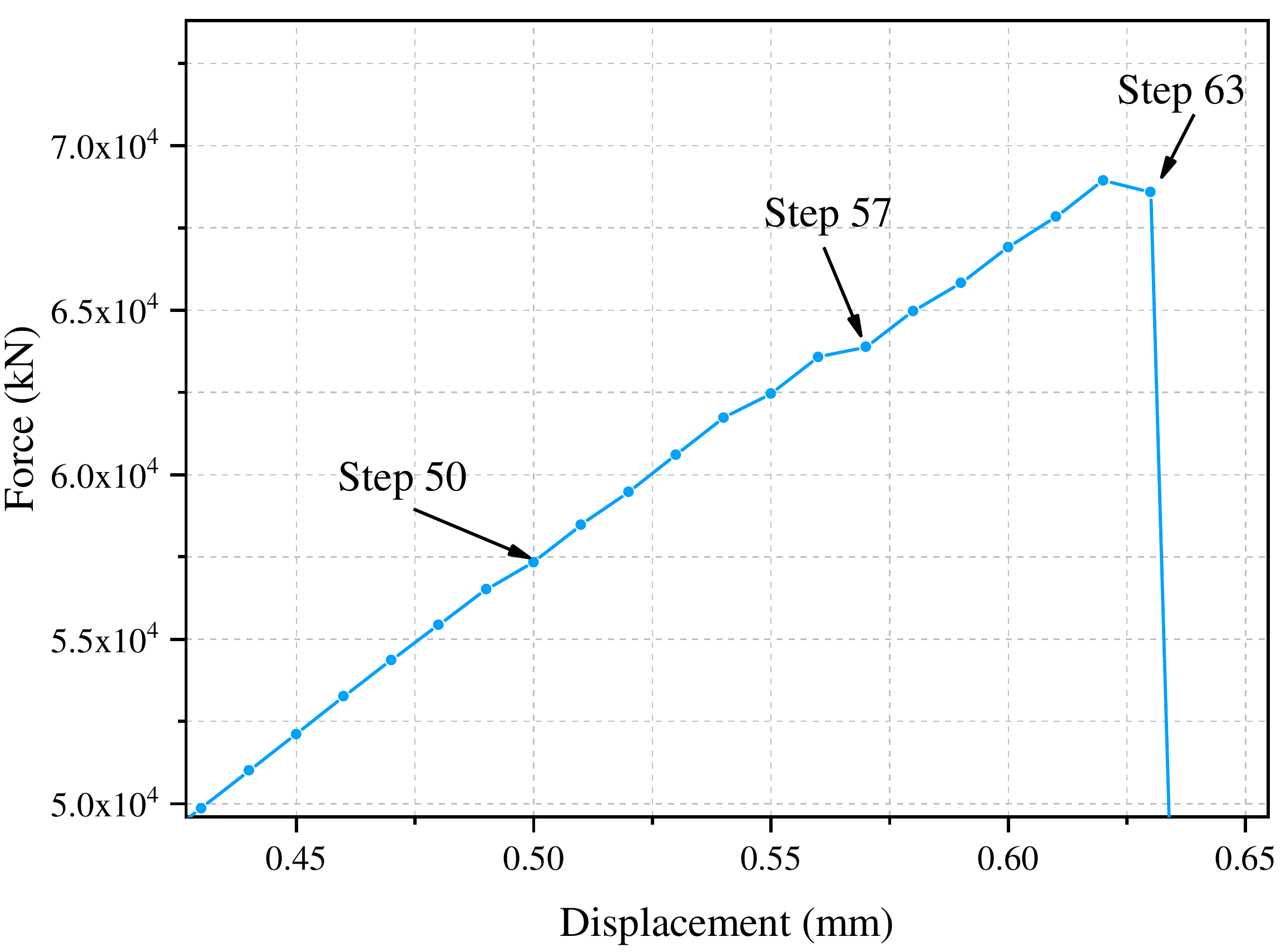}
		\end{minipage}
		\label{2-3b}
	}
	\caption{Force-displacement curves of the Brazilian disk. (a) the full curve; (b) the zooming in of the dotted box in (a).}
	\label{2-3}
\end{figure*}

Fig. \ref{2-3} shows the curves of the average force on the loading points on the top of the disk versus the displacement. Before the effective damage occurs, the curve rises with a constant slope, and then the slope gradually decreases with the increase of breaking bonds. Then, along with the sudden propagation of the cracks between step 63 and step 64, the curve drops drastically.

\subsection{Double-edge-notched plate under tension and shear}
\begin{figure*}[htbp]
	\centering
	\subfloat[]{
		\begin{minipage}[b]{0.5\linewidth}
			\centering
			\includegraphics[width=0.99\linewidth]{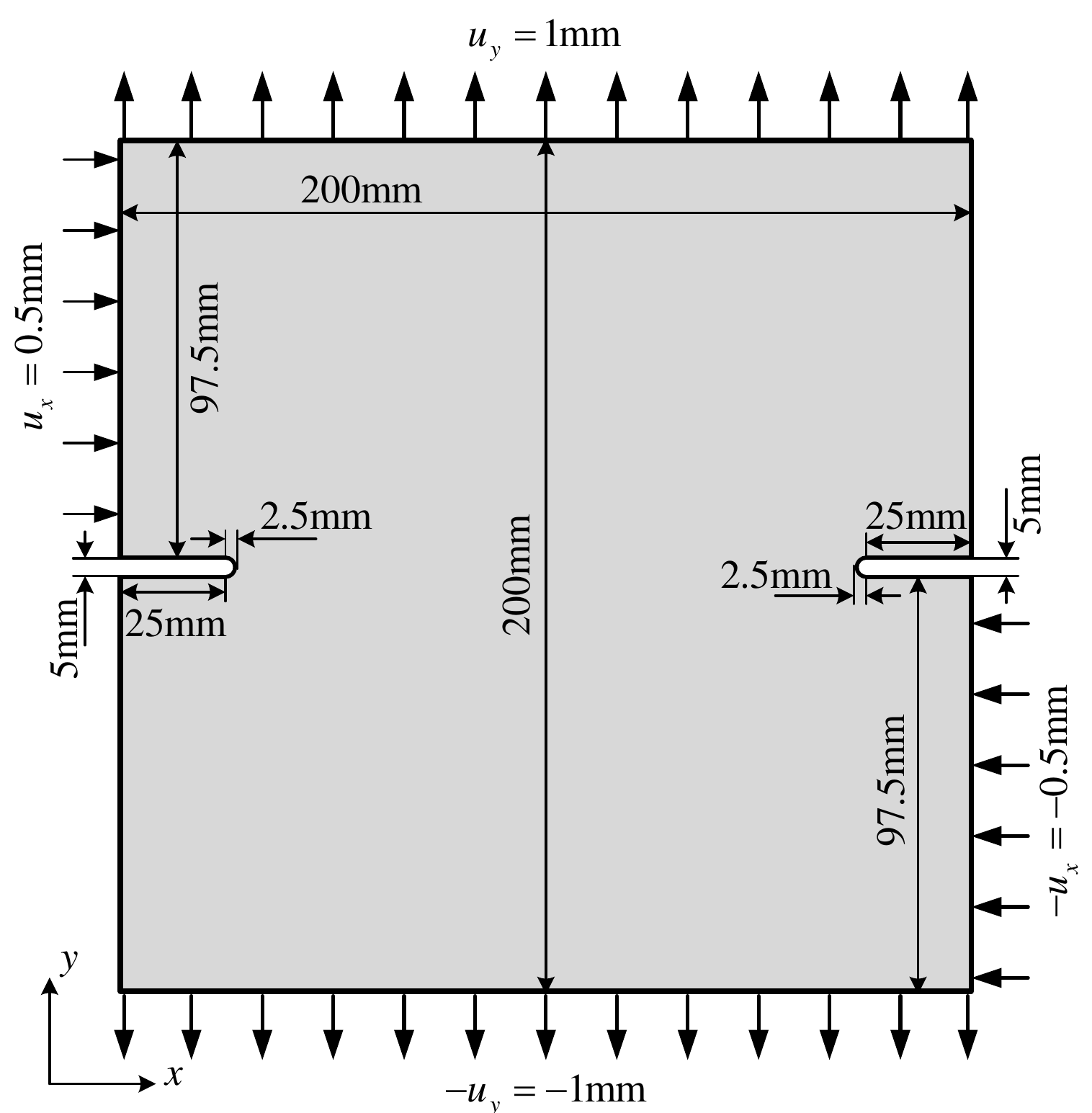}
		\end{minipage}
	\label{3-1a}
	}
	\subfloat[]{
		\begin{minipage}[b]{0.5\linewidth}
			\centering
			\includegraphics[width=0.85\linewidth]{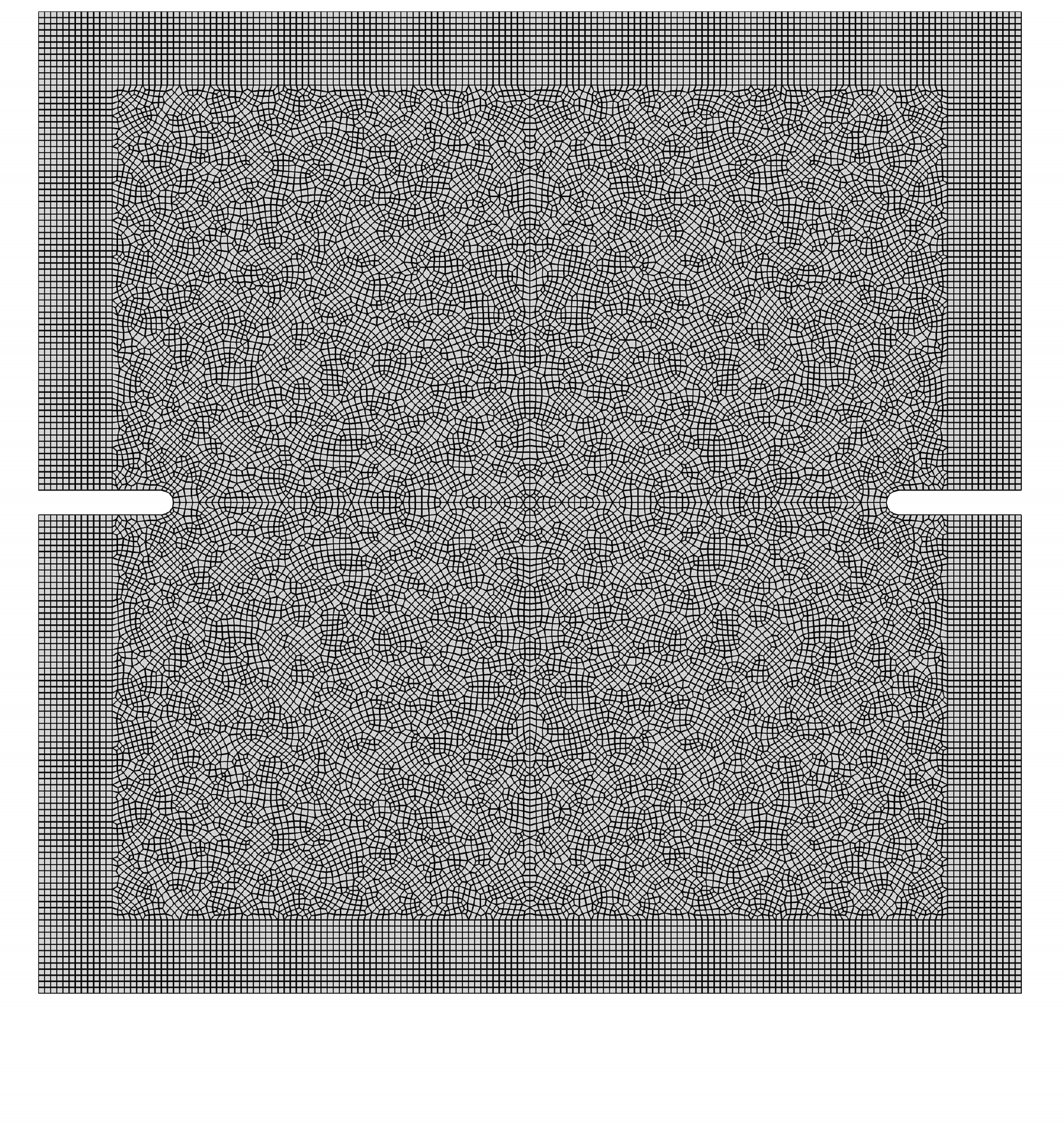}
		\end{minipage}
	\label{3-1b}
	}
	\caption{Schematics for the double-edge-notched plate. (a) the geometry and loading setup; (b) the quadrilateral meshes.}
	\label{3-1}
\end{figure*}

In this example, we consider the mixed mode fracture of a double-edge-notched plate. The geometry and the loading setup of the plate are shown in Fig. \ref{3-1a} and the mesh is shown in Fig. \ref{3-1b}. The Young’s modulus is $E = 30 \text{GPa}$ and the critical stretch is set to be $s_{crit} = 0.02$. The average size of the CEs is $h \approx 1.25 \text{mm}$. The simulation is performed through 25 equably progressive increments steps, i.e., $\text{N}=25$.

\begin{figure*}[htbp] 
	\centering 
	\begin{minipage}[b]{0.98\linewidth} 
		\hfill
		\subfloat{
			\begin{minipage}[b]{0.49\linewidth}
				\centering
				\includegraphics[width=\linewidth]{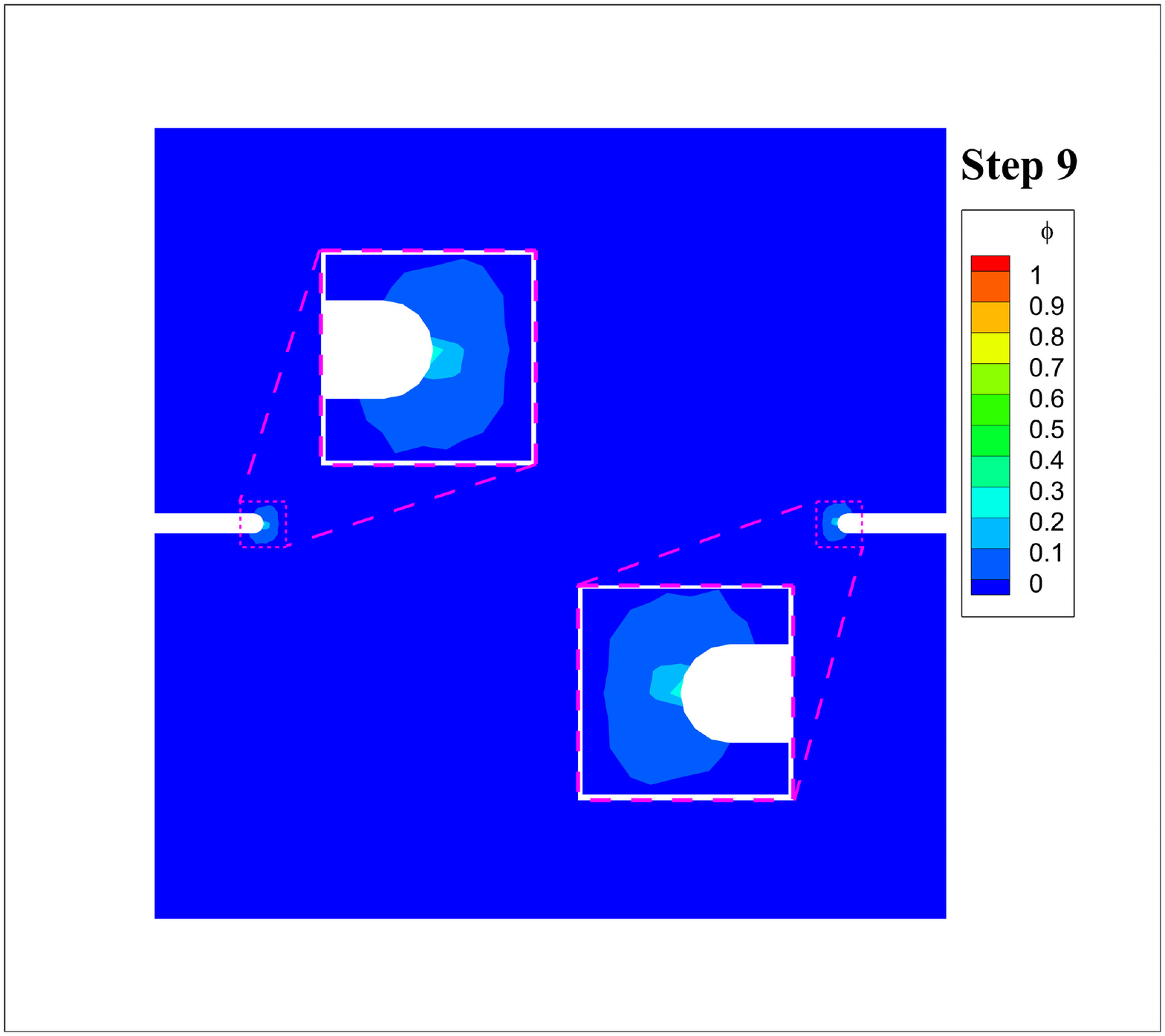}\vspace{8pt}
				\includegraphics[width=\linewidth]{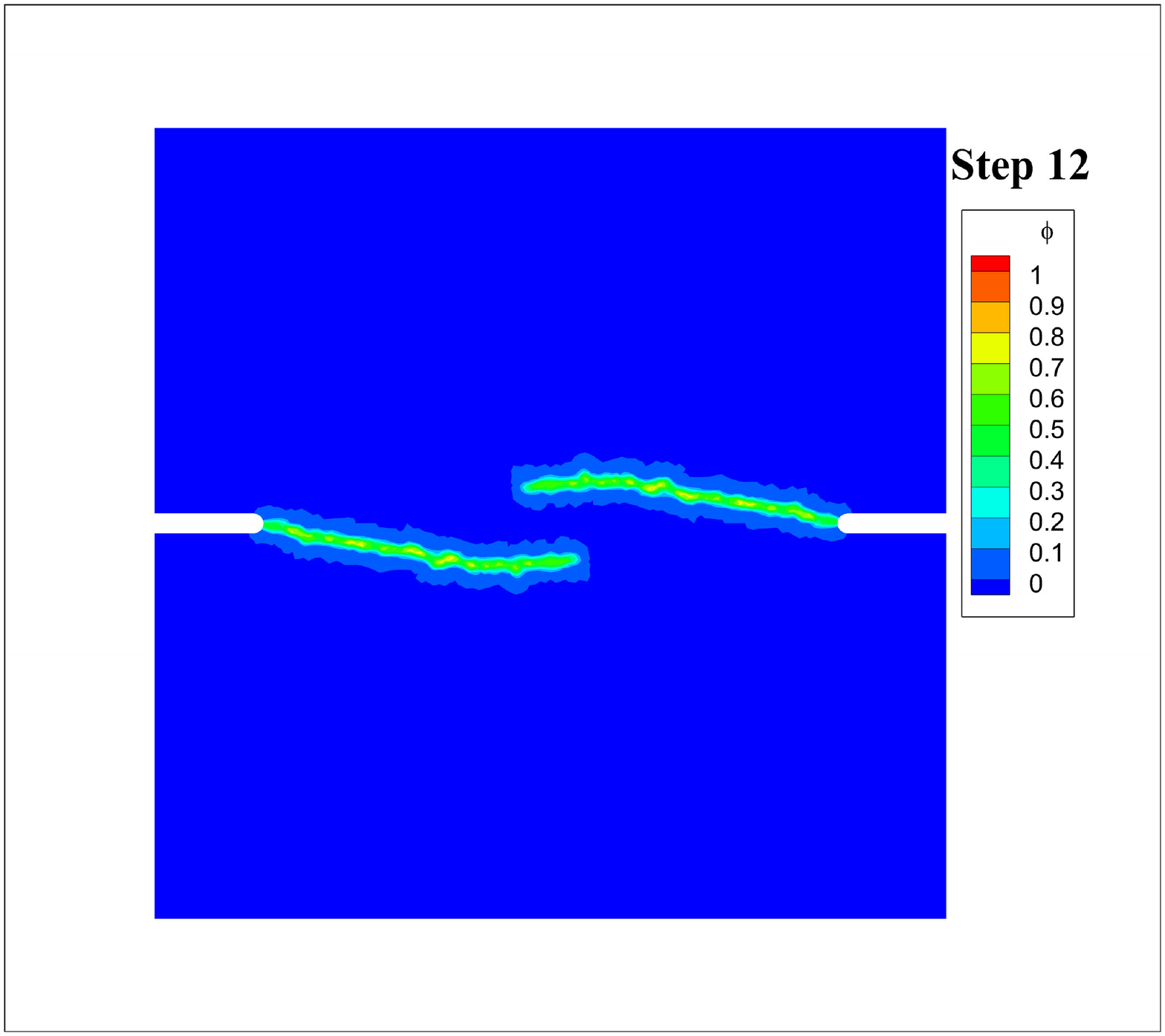}
			\end{minipage}
		}
		%\hfill
		\subfloat{
			\begin{minipage}[b]{0.49\linewidth}
				\centering
				\includegraphics[width=\linewidth]{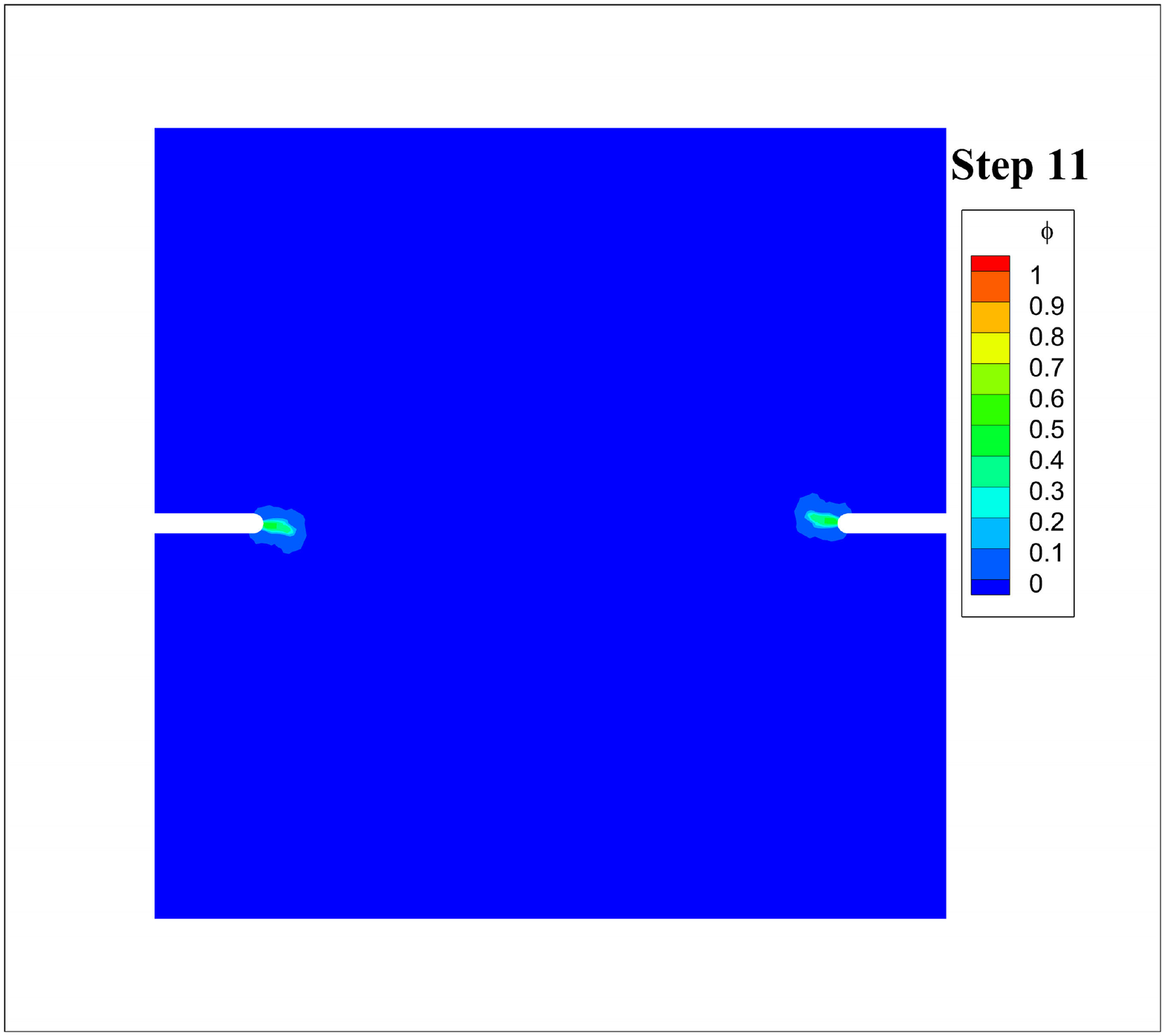}\vspace{8pt}
				\includegraphics[width=\linewidth]{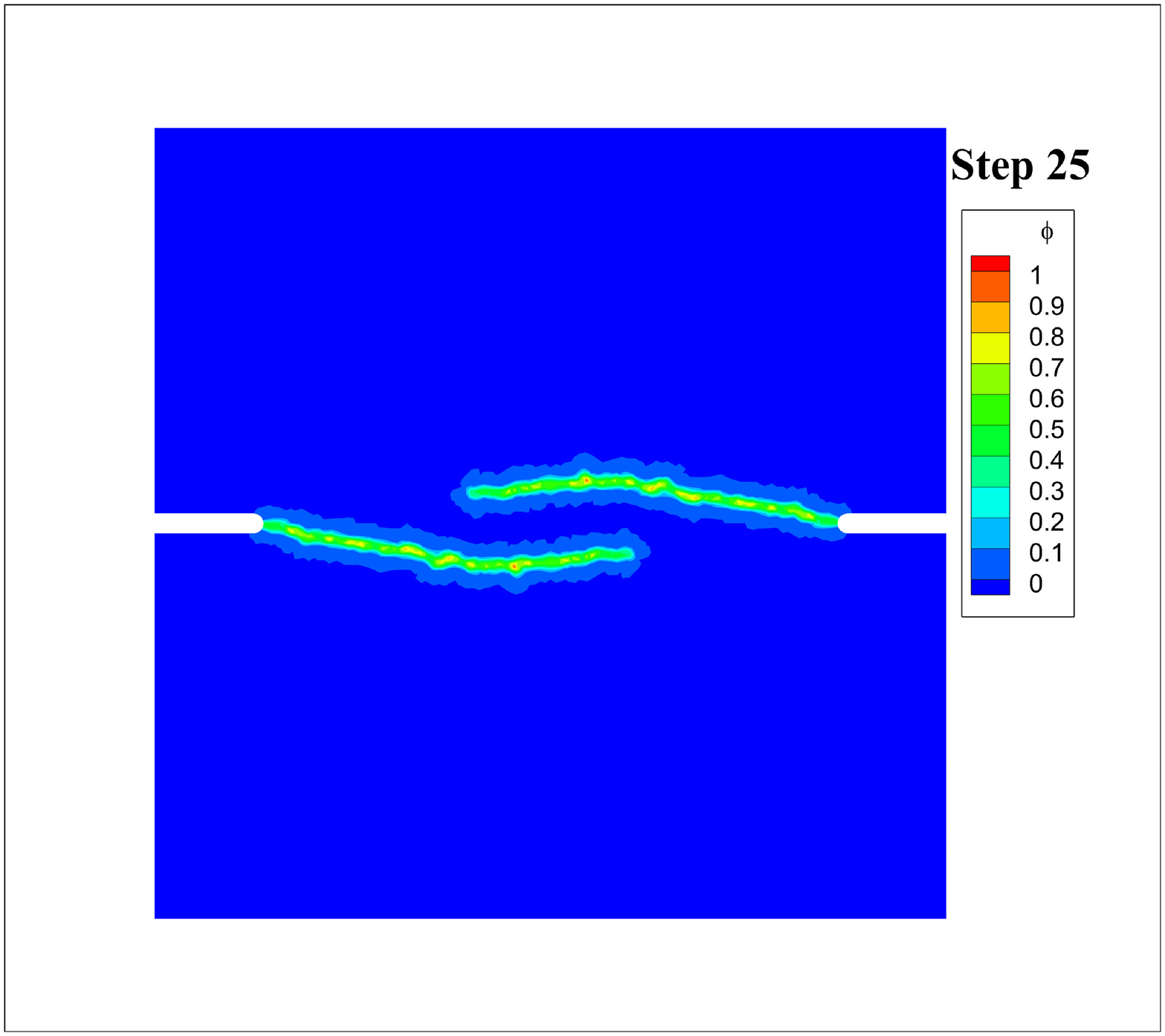}
			\end{minipage}
		}
	\end{minipage}
	\vfill
	\caption{Equivalent damage contours of the double-edge-notched plate under tension and shear.}
	\label{3-2}
\end{figure*}

The effective damage evolution contours of this test are shown in Fig. \ref{3-2}. The damage appears first at the notched corners at step 9. At step 11, the damage is more obvious than at step 9, and it could be found that the damage of the left notched corner propagates down, and the damage of the right notched corner propagates up. Such asymmetric crack path stems from the asymmetry of the boundary conditions, and similar results were reported in \cite[Fig. 17]{18han2016adaptive} and \cite[Fig. 21]{21wang2019hybrid} with the hybrid model. Then at step 12, the cracks propagate destructively.

\begin{figure*}[htbp]
	\centering
	\includegraphics[width=0.55\linewidth]{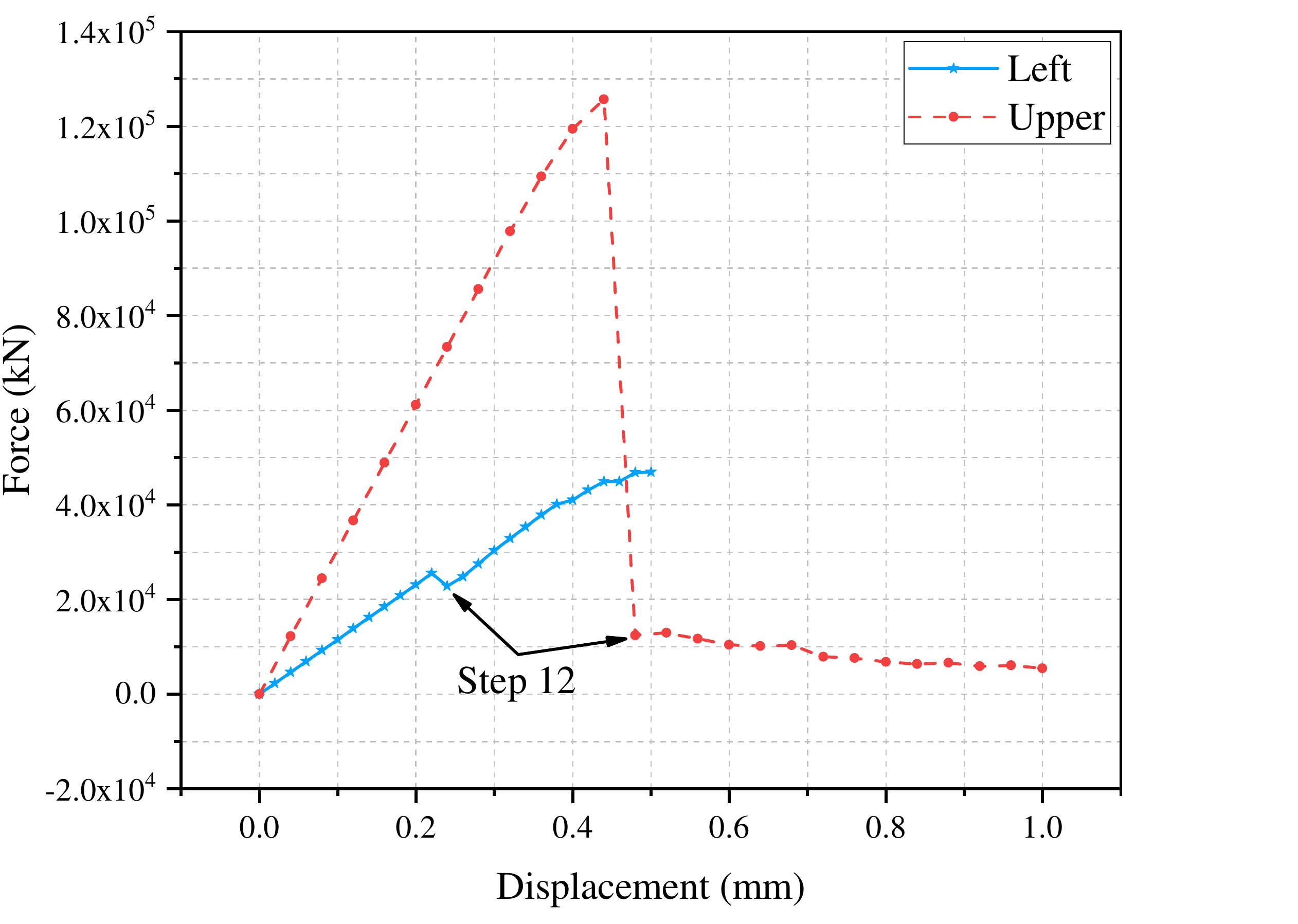}
	\caption{Force-displacement curves of the double-edge-notched plate.}
	\label{3-3}
\end{figure*}

Fig. \ref{3-3} shows the curves of the average force on the loading points on the left and upper edge of the plate versus the displacement. It can be seen that before step 12, the tensile loads on the upper and lower edges of the board play a major role. However, with the destructive propagation of the cracks at step 12, the plate can almost no longer withstand tensile loads. Therefore, the reaction force on the upper edge keeps a very small level after step 12. On the other hand, the force curve related to the left edge keeps rising after a slight drop, which means that the plate can still withstand the shear loads after step 12, and then it is mainly damaged under the shear loads.

%---------------------7---------------------%
\section{Conclusion}\label{S6}
A peridynamics-based finite element method (Peri-FEM) is proposed for the quasi-static fracture analysis in this paper. Three examples were successfully implemented using this method, which demonstrates its feasibility and effectiveness. What is most important is that with the concept of PE, the fundamental computational framework of Peri-FEM is consistent with the classical FEM. Therefore, the numerical algorithm is easy to incorporate with the general FEM software, which will be the focus of our future work.

\section*{Acknowledgment}
The authors gratefully acknowledge the financial support received from the National Natural Science Foundation of China (11872016), Fundamental Research Funds for the Central Universities (DUT20RC(5)005, DUT20LAB203), Key Research and Development Project of Liaoning Province (2020JH2/10500003).

%\section{langrange}
%to the best of our knowledge 

%\section*{References}

\bibliography{mybibfile}

\end{document}